\documentclass[12pt]{article}
\usepackage{amsmath}
\usepackage{amssymb}
\usepackage{graphicx}
\usepackage{geometry}
\usepackage{fancyhdr}
\usepackage{lastpage}
\usepackage{sectsty}
\usepackage[hidelinks]{hyperref}

\sectionfont{\small}
\subsectionfont{\small}

\usepackage[font=small,skip=-6pt]{caption}
\begin{document}
\let\thefootnote\relax\footnotetext{\\[-6pt] $^1$Department of Civil and Environmental Engineering, Tufts University\\$^2$Department of Civil and Resource Engineering, Dalhousie University\\ $^3$Visitor, Earthquake Research Institute, University of Tokyo}
\small

\section*{Abstract}

Models of faults incorporating slip rate- and state-dependent friction have reproduced phenomena from spontaneous slow, aseismic slip to earthquake-generating dynamic rupture. Numerical explorations of model parameter space show sudden transitions in behavior.  However these boundaries are poorly constrained analytically, with commonly used scalings derived assuming unrepresentative conditions of uniform sliding on an infinite, homogeneous fault. In this work, we demonstrate that an analysis of linear stability can reflect model conditions. We examine two scenarios that move beyond the classical case: an asperity driven by the steady creep of its surroundings, and a finite fault experiencing a constant rate of shear loading.  We identify the critical fault dimension $L_c$ at which point linear stability is lost. Beyond this linear regime, the non-linear nature of the friction law implies the loss of memory of loading conditions as instability progresses and the existence of universal solutions describing this process. We refine prior analyses of this non-linear instability and find the minimum fault size that can support self-sustaining, unstable acceleration towards dynamic rupture. We examine the role of the state evolution law and delineate conditions under which faults may be linearly stable but non-linearly unstable, requiring finitely large perturbations to trigger instability. For both the linear and non-linear analysis, we consider scenarios of two- and three-dimensional deformation. On the basis of numerical solutions, approximate but accurate algebraic expressions for the transition boundaries are presented.  These results provide means for careful model design and to delimit plausible regions of parameter space when considering physical observations of stable creep, aseismic (slow slip) or seismic transients.

\section{Introduction}
The development of a slip rate- and state- dependent friction law [{\it Dieterich} 1978, 1979; {\it Ruina}, 1983] was quickly followed by its coupling with elastic deformation in various models to understand the implications for fault phenomena. Elastic interactions were represented either by single- and multiple-degree-of-freedom spring-block models [{\it Gu et al.}, 1984; {\it Rice and Tse}, 1986; {\it Cao and Aki}, 1986; {\it Dieterich}, 1986] or models in which a planar fault is embedded in an elastic continuum. The earliest continuum model accounted for the quasi-static deformation of inter- and pre-seismic stages and a simplification of the dynamic rupture stage provided a complete model of the seismic cycle [{\it Tse and Rice}, 1986]. Subsequent continuum models improved representation of the dynamic phase by incorporating a so-called quasi-dynamic approximation that includes wave-radiation damping effects but neglects wave-mediated interactions between points on the fault [{\it Rice}, 1993; {\it Ben-Zion and Rice}, 1995; {\it Kato and Hirasawa}, 1997]. Such quasi-dynamic models continue to be frequently used due to computational convenience. More computationally costly rate-and-state models with complete elastodynamic response followed in two [{\it Lapusta et al.}, 2000] and three dimensions [{\it Lapusta and Liu}, 2009]. While often showing qualitative similarities with simpler quasi-dynamic models, there are also several fundamental differences apparent during dynamic rupture [e.g., as documented by {\it Lambert and Lapusta}, 2021]. 

\

These early models were concerned with earthquake sizes, distributions, and recurrence; however, following the geodetic observation of slow-slip events [{\it Hirose et al.}, 1999; {\it Dragert et al.}, 2001], it became apparent that periods of transient, aseismic acceleration of the fault could also emerge from continuum models of faults with rate- and state-dependent friction [{\it Kato}, 2003; {\it Liu and Rice}, 2005; {\it Shibazaki and Shimamoto}, 2007]. Moreover, parametric studies indicated that aseismic transients without dynamic rupture occur in regions of parametric space bounded by regions for which either stable creep or dynamic rupture occur [{\it Kato}, 2003; {\it Liu and Rice}, 2007; {\it Rubin, 2008}]. Subsequent work has more extensively explored this parameter space, and though a rich variety of behaviors are apparent, there still remain distinct boundaries between steady creep, transient slow slip, and dynamic rupture [e.g., {\it Barbot}, 2019; {\it Veedu et al.}, 2020; {\it Nie and Barbot}, 2021; {\it Zhai and Huang}, 2024]. 

\

These continuum models typically consist of a seismogenic region along the fault, in which properties exhibit a tendency to weaken with increasing slip rate and hence potentially susceptible to earthquake-nucleating instability, regions that strengthen with increasing slip rate and where slip may be stably accommodated, and a mechanical representation of far-field tectonic loading. In all of the continuum models, a fundamental concern is how large must the rate-weakening region be to give rise to an earthquake or aseismic transient and what length scale must be resolved by numerical solutions [e.g., {\it Ben-Zion and Rice}, 1995]. An elasto-frictional lengthscale determined by early linear stability analysis [{\it Ruina}, 1983; {\it Rice and Ruina}, 1983] has provided an approximate measure of susceptibility of a rate-weakening region to instability [{\it Rice}, 1993; {\it Liu and Rice}, 2007; {\it Rubin}, 2008]. We take a closer look at this linear stability analysis and its connection with fault models.

\

{\it Ruina} [1983] introduced a general rate- and state-dependent formulation for sliding friction, accompanied by a linear stability analysis of a spring-block model, identifying a critical threshold stiffness below which steady sliding is unstable to perturbations. {\it Rice and Ruina} [1983] utilized a mapping between perturbations to a spring-block system and periodic perturbations along a frictional interface within a continuum, considering the specific cases of in-plane (mode-II) or out-of-plane (mode-III) rupture of a fault plane dividing two elastic half-spaces or slabs of fixed and equal height. In either case, the fault extends infinitely along the two directions of its plane and the perturbations are about uniform sliding at steady state; the mapping of {\it Rice and Ruina} [1983] translated the critical stiffness of {\it Ruina} [1983] to a critical spatial perturbation wavelength above which perturbations may grow unstably. While this analysis provided an early indication of the stability of model faults and the critical wavelength has been a canonical reference for a nucleation length [e.g., {\it Segall}, 2010], the basis of this continuum analysis is not representative of the wide variety of model conditions and neglects the non-linear nature of the friction law. Seismic cycle models typically have variations in frictional properties along down-dip or along-strike directions; the fault may be subject to an increasing stress due to loading conditions; and the fault typically has a finite domain, with its boundary either satisfying a locked condition, a prescribed slip rate intended to model tectonic forcing, or a free-surface condition. While possibly providing an initial indication of model stability, periodic perturbations along an unbounded fault cannot capture this range of properties, loading, and boundary conditions.

\

\begin{figure}[t]
      \center\includegraphics[width=263pt]{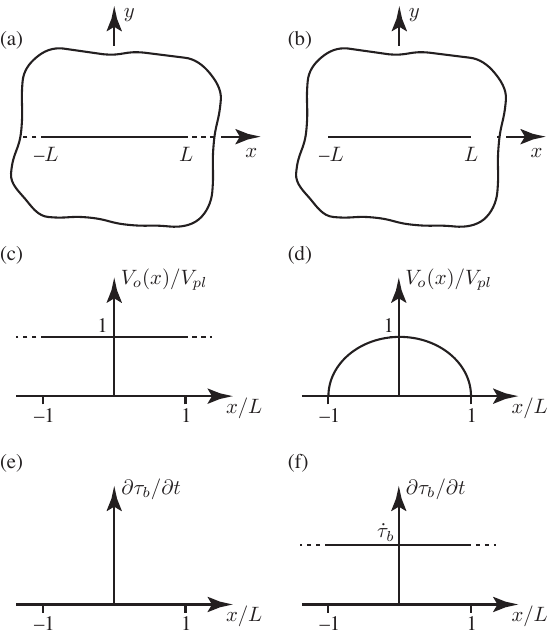}\vspace{5 mm}
   \caption{\raggedright Proceeding down the left and right columns of panels, we compare two elementary examples of faults with finite-sized rate-weakening regions, their steady-state sliding rates $V_o(x)$, and shear stress loading rates. (a) Schematic of finite-sized rate-weakening asperity of length $2L$ embedded within a much longer fault that creeps at a constant rate $V_{pl}$. (b) Schematic profile of fault of length $2L$ that is loaded by a constant rate of shear stress  $\dot\tau_b$ (not illustrated). The fault may terminate at $x=\pm L$ or the fault may continue along $|x|>L$ but is locked there (i.e., slip is prohibited). (c,d) Plot of steady-state distribution of slip rate, $V_o(x)$ for scenarios (a) and (b), respectively, scaled by a characteristic slip rate from loading, $V_{pl}$. (e,f) Plot of the rate of ``background" shear stress, $\partial \tau_b/\partial t$, defined as the rate of shear stress on the fault in the absence of non-uniform slip, for the scenarios (a) and (b), respectively. For (e), $\partial\tau_b/\partial_t=0$.}
   \label{fig:schem}
\end{figure}

To address this first deficiency, in section \ref{sec:linstab} of this work, we revisit the problem of linear stability and consider two elementary cases that exhibit the effects of fault finiteness and different loading configurations (Figure \ref{fig:schem}). First, we consider a finite-sized asperity sliding at steady state while surrounded by sliding that remains at a fixed rate. Second, we examine a finite fault whose boundaries are locked and the fault loaded by a uniform, constant rate of shear stress. As also discussed by {\it Ciardo and Viesca} [2025], we show that the first case can be considered by the same mapping to a spring-block system pursued by {\it Rice and Ruina} [1983]. The second case requires special treatment owing to the non-uniformity of slip rate and state, about which linear perturbations are considered. In both cases we identify a critical value of a dimensionless parameter, above which the fault is linearly unstable. This value can be interpreted as identifying a critical size of the rate-weakening fault or the rate-weakening patch on a fault. Notably we find that inherent non-uniformity of fault slip on finite faults leads to a new, distinct scaling of the critical fault size with the friction rate-weakening parameter, $b-a$, suggesting that the classical critical length for uniform slip on unbounded faults can grossly underpredict the critical dimension of finite faults.

\

To address the second deficiency of neglecting frictional non-linearity, in section \ref{sec:nonlin} we revisit and refine prior analyses in which solutions for self-sustaining non-linear instability are found [{\it Rubin and Ampuero, 2005}; {\it Viesca}, 2016a,b; {\it Viesca}, 2023]. In contrast with the linear stability analysis, these solutions describe quasi-static acceleration towards dynamic rupture taking into account the full non-linearity of rate-and-state friction. A remarkable feature of these solutions is their near-independence of the loading and boundary conditions, or universality. We highlight one exception to this independence and differentiate between so-called ``free" and ``pinned" solutions: the former qualification applies when faults are sufficiently large that this non-linear instability can develop without encountering any locked, steadily creeping, or terminal boundaries of a fault, and the latter applies to faults that are sufficiently small to constrain the lateral extent of instability development. These lengthscales can be used to identify a critical nucleation size for dynamic rupture, distinct from the critical dimension identified by linear stability analysis.

\

In the subsequent sections, we use the results of linear and non-linear instability development to make a priori predictions of fault model behavior on the basis of the choice of fault frictional properties, rock elastic properties, fault dimensions and loading. In section \ref{sec:trans}, we highlight the relevance of our results to understand the transition from linear to non-linear stages of instability development. In section \ref{sec:disc} we discuss in greater detail several aspects of the results. Foremost are the implications for quantitatively and precisely predicting the behavior of seismic cycle models. Additionally, we will show extensions of the results of sections \ref{sec:linstab} and \ref{sec:nonlin}, which consider the two-dimensional elastic response to a single-mode of slip with one-dimensional variations along the fault, to three-dimensional elasticity, involving mixed-mode slip with variations occurring across two dimensions of the fault; and finally, we will discuss consequences of changing the so-called state-evolution law. The techniques are evidently generalizable to more complex cases, such as spatially varying frictional properties and effective normal stress.

\section{Governing equations and linearization about steady state}
\label{sec:ge}

We consider a planar fault lying in the $x$-$y$ plane and sliding such that displacement varies within the $x$-$z$ plane either as an in-plane or anti-plane state of deformation. This implies that the fault-resolved shear stress can be written as 
\begin{equation}
\tau(x,t)=\tau_b(x,t)+\mathcal{L}[\delta(x,t)]
\label{eq:t}
\end{equation}
where $\delta$ is the distribution of relative displacement across the fault plane, or simply, slip, $\tau_b$ is the fault-resolved shear stress in the absence of slip, and $\mathcal{L}$ is a linear operator that determines the instantaneous, quasi-static change in shear stress along the fault due to a distribution of slip. The operator has units of stress per unit distance. In the example of a fault dividing two elastic half-spaces [e.g., {\it Rice}, 1968],
\begin{equation}
\mathcal{L}[\delta(x,t)]=\frac{\mu'}{\pi}\int_{-\infty}^\infty \frac{\partial\delta(\xi,t)/\partial \xi}{\xi-x}d\xi
\end{equation}
where $\mu'=\mu/[2(1-\nu)]$ and $\mu'=\mu/2$ for in-plane and anti-plane deformation.

\

The shear strength is $\tau_s(x,t)=\sigma f(x,t)$ where $\sigma$ is the fault-normal stress and $f$ is a rate- and state-dependent friction coefficient [{\it Ruina}, 1983] that may be written as 
\begin{equation}
f(x,t)=f_i+a \ln\left(\frac{V(x,t)}{V_i}\right)+b\ln\left(\frac{\theta(x,t)}{\theta_i}\right)
\label{eq:f}
\end{equation}
where $V=\partial\delta/\partial t$ is the slip rate, $\theta$ is a state variable, $f_i$, $V_i$, and $\theta_i$ are reference values for friction, sliding and state, and the coefficients $a$ and $b$ are dimensionless parameters. $a$ is referred to as the direct-effect coefficient and is a pre-factor of the direct dependence of $f$ on slip rate. $b$ is referred to as the evolution-effect coefficient and pre-factor of the dependence of $f$ on the state variable. Both coefficients are empirically determined. A number of evolution laws were proposed for the state variable, [{\it Ruina}, 1983]
\begin{subequations}
\begin{align}
\frac{\partial\theta}{\partial t}&= 1-\frac{V \theta}{D_c}\quad \text{(aging law)} \label{eq:ag}\\[4 pt]
\frac{\partial\theta}{\partial t}&=-\frac{V\theta}{D_c}\ln\frac{V\theta}{D_c} \quad \text{(slip law)} \label{eq:sl}\\[4 pt]
\frac{\partial\theta}{\partial t}&=\frac{1}{\epsilon}\left[\left(\frac{V\theta}{D_c}\right)^{-\epsilon}-1\right]\frac{V\theta}{D_c} \quad \text{(intermediate law)} \label{eq:in}\end{align}
\label{eq:ts}
\end{subequations}
The last evolution law, (\ref{eq:in}), corresponds to a one-parameter family of laws in terms of $\epsilon$, which allows to continuously transition from the aging law ($\epsilon=1$) to the slip law ($\epsilon\rightarrow0$) [{\it Ruina}, 1983; {\it Viesca}, 2023]. Each law is a function of the dimensionless parameter $V\theta/D_c$. $D_c$ is a characteristic slip distance over which the memory of recent sliding history is lost [{\it Ruina}, 1983]. The three laws share the steady-state point  $V\theta/D_c=1$, when $\partial\theta/\partial t=0$, and a common linearization about this point. The steady-state value of friction $f_{ss}$ depends logarithmically on the slip rate
\begin{equation}
f_{ss}(V)=f_i+(a-b)\ln(V/V_i)
\end{equation}
where the difference of the direct and evolution effect coefficients determines whether the steady-state behavior is rate-weakening or rate-strengthening. We will be concerned with the emergence of instabilities stemming from rate-weakening behavior, when $a<b$. Furthermore, well-posedness requires $a>0$.

\

We now look to linearize the pair of non-linear evolution equations for $V$ and $\theta$ that result from the combination of (\ref{eq:t}), (\ref{eq:f}), and one of (\ref{eq:ts}). We will be concerned with the evolution of departures $\tilde V$ and $\tilde \theta$ from a regime of steady-state sliding under slip rate $V_o$ and state $\theta_o$ 
\begin{subequations}
\begin{align}
V(x,t)&=V_o(x)+\tilde V(x,t)\\
\theta(x,t)&=\theta_o(x)+\tilde \theta(x,t)
\end{align}
\label{eq:per0}
\end{subequations}
$V_o$ and $\theta_o$ are not necessarily uniform in space but are such that 
\begin{equation}
\frac{V_o(x)\theta_o(x)}{D_c}=1
\end{equation} The three state evolution laws hence linearize to 
\begin{equation}
\frac{\partial \tilde\theta}{\partial t}=-\frac{\tilde\theta(x,t)}{\theta_o(x)}-\frac{\tilde V(x,t)}{V_o(x)}
\label{eq:lin1}
\end{equation}
and the friction evolution equation linearizes to
\begin{equation}
\frac{\partial f}{\partial t}=\frac{a}{V_o(x)}\frac{\partial \tilde V}{\partial t}+\frac{b}{\theta_o(x)}\frac{\partial \tilde\theta}{\partial t}
\end{equation}
Also requiring that $\partial\tau/\partial t=\sigma\partial f/\partial t$, implies that, to within linearization,
\begin{equation}
\mathcal{L}[\tilde V(x,t)]=\sigma\left(\frac{a}{V_o(x)}\frac{\partial \tilde V}{\partial t}+\frac{b}{\theta_o(x)}\frac{\partial \tilde\theta}{\partial t}\right)
\label{eq:lin2}
\end{equation}
and that $V_o$ and $\partial \tau_b/\partial t$ must satisfy
\begin{equation}
\frac{\partial \tau_b}{\partial t}+\mathcal{L}[V_o(x)]=0
\label{eq:tb}
\end{equation}
The latter requirement follows from considering that the unperturbed regime is steady such that both state $\theta_o$ and slip rate $V_o$ do not evolve.

\section{Linear stability of spatially uniform and non-uniform steady states of sliding}
\label{sec:linstab}

In this section we analyze the linear stability of faults under uniform and non-uniform steady-state sliding. In section \ref{sec:uni}, we consider the stability of uniform, steady sliding of an unbounded fault (section \ref{sec:infin}) and of a finite-sized rate-weakening asperity loaded by surrounding creep (section \ref{sec:fin}). We highlight that critical dimensions for the former and latter cases (a perturbation wavelength and asperity size, respectively), exhibit the same, classical scaling with the steady-state friction rate-weakening parameter, $(b-a)^{-1}$. In section \ref{sec:finuni}, we examine a distinct problem in which a finite fault is loaded by a constant stressing rate and for which the steady-state solution for slip rate is non-uniform. We demonstrate that this non-uniformity leads to a change of scaling of the critical fault size at which point stability is lost, such that it asymptotically scales as $(b-a)^{-3/2}$ in the limit $a\rightarrow b$. The finite-asperity and finite-fault cases of sections \ref{sec:fin} and \ref{sec:finuni} can be considered as end-members of fault loading, which represent well-defined physical scenarios.

\subsection{Faults with initially steady, uniform sliding rate}
\label{sec:uni}
Revisiting the linear stability analysis of {\it Rice and Ruina} [1983], except here for a generic form of elastic stress transfer operator $\mathcal{L}$, we consider a fault for which $\partial\tau_b/\partial t=0$ and we perturb about uniform values of slip rate and state 
\begin{equation} V_o(x)=V_{pl}, \quad \theta_o(x)=D_c/V_{pl} \label{eq:Vo1}\end{equation}
We pass from dimensional to non-dimensional variables in the manner
\begin{equation}
\frac{V}{V_{pl}}\Rightarrow V, \quad \frac{\theta,\,t}{D_c/V_{pl}}\Rightarrow \theta, t\,, \quad \frac{D_c \mathcal{L}}{\sigma b}\Rightarrow \mathcal{L}
\label{eq:nd}
\end{equation}
such that the perturbations $\tilde V$ and $\theta$ are similarly non-dimensional and the linearized governing equations are
\begin{subequations}
\begin{align}
\mathcal{L}(\tilde V)=\frac{a}{b} \frac{\partial \tilde V}{\partial t} +\frac{\partial \tilde\theta}{\partial t}\\[9 pt]
\frac{\partial \tilde \theta}{\partial t} = -\tilde V -\tilde\theta
\end{align}
\end{subequations}
We look for perturbations in the form
\begin{equation}
\tilde V(x,t) = \mathcal{V}(x) e^{st} ,\quad \tilde \theta(x,t)=\Theta(x) e^{st} 
\label{eq:pert}
\end{equation}
where the exponent $s$ and spatial distribution $\mathcal{V}$ and $\Theta$ of the perturbation are to be determined. The exponent $s$ may be complex. Its real component gives the rate of exponential growth or decay and its complex component gives a frequency of harmonic oscillation. The governing equations reduce to
\begin{subequations}
\begin{align}
\mathcal{L}(\mathcal{V})=\frac{a}{b}s\mathcal{V}+s\Theta\\[9 pt]
s\Theta = -\mathcal{V} -\Theta
\end{align}
\label{eq:fsys0}
\end{subequations}
the second of which allows us to eliminate $\Theta$ from the first since $\Theta=-\mathcal{V}/(1+s)$, leaving us with 
\begin{equation}
\mathcal{L}(\mathcal{V})=\left(\frac{a}{b}s-\frac{s}{1+s}\right) \mathcal{V}
\label{eq:eig0}
\end{equation}
which has the form of an eigenvalue problem where we identify the terms in parentheses on the right hand side as the eigenvalue $-k$, such that
\begin{equation}
\mathcal{L}(\mathcal{V})=-k \mathcal{V}
\label{eq:eig}
\end{equation}
and
\begin{equation}
0=\frac{a}{b}s^2+s(k+a/b-1)+k
\end{equation}
and $k$ bearing the meaning of the non-dimensional fault stiffness in response to a spatial mode $\mathcal{V}$.
We may solve the quadratic equation for $s$ in terms of $k$ and $a/b$
\begin{equation}
s=\frac{(1-a/b-k)\pm\sqrt{(1-a/b-k)^2-4ka/b} }{2 a/b}
\label{eq:s}
\end{equation}
and identify the occurrence of a Hopf bifurcation, when $s$ takes on a complex conjugate pair of values that cross the imaginary axis when the term in parentheses is zero. This defines a critical value for $k$
\begin{equation}
k_c=1-a/b
\end{equation}
which can be considered as a dimensionless critical stiffness for the system, below which the real part of the growth exponent $s$, (\ref{eq:s}), is positive and perturbations are unstable. The corresponding conjugate values of $s$ are
\begin{equation}
s_c=\pm i \sqrt{b/a-1}
\label{eq:sc1}
\end{equation}

\

As will be illustrated in the following two subsections, the eigenvalue problem (\ref{eq:eig}) is pivotal and its solution enables the translation of the critical value $k_c$ to a critical spatial dimension along the fault. That translation is determined by the particular form of the elastic stress-transfer operator $\mathcal{L}$, which reflects the fault geometry, mode of sliding, and presence of free surfaces. {\it Ciardo and Viesca} [2025] recognized this in the context of a study of spring-block slider stability and solved this eigenvalue problem analytically and numerically for a number of elementary configurations, including those presented here. (The reader should be aware of a typographical error of an omitted minus sign in their equation (1.8): there is no consequence of this omission except that $\mathcal{L}$ therein is minus that considered here.)

\subsubsection{Infinite fault}
\label{sec:infin}
We use the above to rederive the results for the specific case considered by {\it Rice and Ruina} [1983; see also {\it Segall}, 2010] of an unbounded fault, for which the dimensionless expression of the operator $\mathcal{L}$ is
\begin{equation}
\mathcal{L}(V)=\frac{D_c \mu'}{\sigma b}\frac{1}{\pi} \int_{-\infty}^\infty \frac{\mathcal V'(\xi)}{\xi-x}d\xi
\end{equation}
where $\mu'=\mu/2$ for anti-plane (mode-III) sliding and $\mu'=\mu/[2(1-\nu)]$ for in-plane (mode-II) slip. We non-dimensionalize distance as 
\begin{equation}  \quad x \frac{\sigma b}{D_c\mu'}\Rightarrow x\end{equation}
such that the eigenvalue problem (\ref{eq:eig}) is
\begin{equation} 
\frac{1}{\pi} \int_{-\infty}^\infty \frac{\mathcal{V}'(\xi)}{\xi-x}d\xi=-k \mathcal V
\label{eq:eig1}
\end{equation}
We identify the left hand side of (\ref{eq:eig1}) as the negative of the Hilbert transform $\mathcal{H}(f)=\frac{1}{\pi}\int_{-\infty}^\infty \frac{f(\xi)}{x-\xi} d\xi$ and can write for brevity
\begin{equation}
\mathcal{H}(\mathcal{V'})=k \mathcal{V}
\label{eq:eig1a}
\end{equation}
The Hilbert transform and the spatial derivative have the commutative property $(\mathcal{H}(f))'=\mathcal{H}(f')$, provided $f(x)$ decays sufficiently fast as $x\rightarrow\pm\infty$ [e.g., {\it King}, 2009]. In addition $\mathcal{H}$ has the property of being its own inverse, to within a sign change: $\mathcal{H}[\mathcal{H}(f)]=-f$. Keeping these two points in mind, we can reduce (\ref{eq:eig1a}) to [{\it Ciardo and Viesca}, 2025]
\begin{equation}
\mathcal{V}''+k^2\mathcal{V} = 0 
\end{equation}
which has the two linearly independent solutions $\cos(kx)$ and $\sin(kx)$ and we find that $k$ corresponds to a wavenumber, or a perturbation wavelength $\lambda$ with $k=2\pi/\lambda$. The critical value $k_c$ in this case hence implies a corresponding critical wavelength $\lambda_c$, above which perturbations are unstable,
\begin{equation}
\lambda_c=\frac{2\pi}{1-a/b}
\end{equation}
Reverting to dimensional form for the critical wavelength via $\lambda_c\Rightarrow \lambda_c \,\sigma b/(\mu' D_c)$ and using the lengthscale \begin{equation}L_b=2\mu' D_c/(\sigma b)
\label{eq:Lb}
\end{equation}
 introduced by {\it Rubin and Ampuero} [2005], $\lambda_c$ can be expressed as
\begin{equation}
\lambda_c=\frac{\pi L_b}{1-a/b}
\label{eq:lamc}
\end{equation}
or more simply,
\begin{equation}
\lambda_c=\pi L_{b-a}
\end{equation}
where we use an alternative lengthscale $L_{b-a}=2\mu' D_c/[\sigma(b-a)]$, also introduced by {\it Rubin and Ampuero} [2005]. 

\

Subsequent works have examined linear stability of unbounded faults with rate- and state-dependent friction in conjunction with other physical processes arising from material [{\it Rice et al.}, 2001] or geometric [{\it Aldam et al.}, 2016; {\it Brener et al.}, 2016] dissimilarity, or effects of dilatancy, compaction, poroelasticity, and fault permeability evolution [{\it Segall and Rice}, 1995; {\it Heimisson et al.}, 2019, 2021; {\it Ozawa et al.}, 2024; and {\it Mei}, 2025]. All of these works consider the evolution of linearized equations and identify critical stiffnesses, or equivalent critical wavelengths for harmonic perturbations, for loss of stability, paralleling the analysis of {\it Rice and Ruina} [1983].

\subsubsection{Finite asperity}
\label{sec:fin}
The results of the preceding subsection establishing a critical wavelength for an unbounded fault are known and were established by {\it Rice and Ruina} [1983]. We now look to examine a novel case, in which a finite rate-weakening region $|x|<L$ is surrounded by a region $|x|\geq L$ that steadily slides at a rate $V_{pl}$: This configuration may be thought of as a creeping region that encloses an asperity of size $2L$, for which we are interested in the response to perturbations about steady-state sliding (\ref{eq:Vo1}) within $|x|<L$. For this case the dimensionless form of $\mathcal{L}$ is
\begin{equation}
\mathcal{L}(V)=\frac{D_c\mu'}{\sigma b}\frac{1}{\pi}\int_{-L}^L\frac{V'(s)}{s-x}ds
\label{eq:fin}
\end{equation}
and we may subsequently non-dimensionalize distance as $x/L\Rightarrow x$ such that the eigenvalue problem (\ref{eq:eig}) becomes
\begin{equation}
\frac{1}{2\pi} \int_{-1}^1 \frac{\mathcal{V}'(s)}{x-s}ds = k \frac{L}{L_b} \mathcal{V}(x)
\label{eq:eig2}
\end{equation}
in which $L$ remains dimensional and the ratio $L/L_b$ can be considered as a dimensionless fault half-length. By (\ref{eq:per0}) and (\ref{eq:pert}), the boundary condition $V(\pm L,t)=V_o$, implicitly imposes the condition on the perturbation that $\mathcal{V}(\pm 1)=0$. We recognize (\ref{eq:eig2}), subject to this boundary condition, as the eigenvalue problem examined by {\it Uenishi and Rice} [2003], for which the combination $k L/L_b$ corresponds to the eigenvalue $\lambda$ in that work. This eigenvalue has a discrete spectrum with a countably infinite number of values, the lowest three of which are $(kL/L_b)_1=0.57888694...$, $(k L/L_b)_2=1.3773773...$, and $(k L/L_b)_3=2.1584005...$ (see Appendix \ref{app:URsoln} for numerical solution). To determine the critical value of $L$ above which the fault is linearly unstable, $L_c$, the salient eigenvalue is the smallest one. Since linear instability is incipient when $k=k_c$, $L_c$ can be expressed as
\begin{equation}
L_c= \frac{(0.578...)}{1-a/b}L_b\quad \text{or,}\quad L_c= (0.578...)L_{b-a}
\label{eq:Lc1}
\end{equation}
A depiction of the eigenfunctions is available in {\it Uenishi and Rice} [2003]. The eigenfunction corresponding to the leading eigenvalue can be simply and accurately approximated as [{\it Viesca and Dublanchet}, 2019]
\begin{equation}
\mathcal{V}_1(x)\approx\sqrt{1-x^2}(1-x^2/3)
\label{eq:eigapprox}
\end{equation}
We note that the critical distances for both cases of uniform sliding, $L_c$ and $\lambda_c$, scale with $L_{b-a}$. Comparing the total critical asperity length $2L_c$ with the inter-nodal distance $\lambda_c/2$, yields $4L_c/\lambda_c =(1.157...)/(\pi/2)=0.737...$. Thus the implicit use by the modeling community of $\lambda_c$ in estimating critical fault dimensions appears to be a justifiable, if imprecise, calculation. As we will observe in the following section \ref{sec:finuni}, this expectation turns out to be wrong when the loading is such that the baseline sliding is strongly non-uniform.

\subsection{Finite faults with initially steady, non-uniform sliding rate}
\label{sec:finuni}
A non-uniform sliding rate is an inherent feature of faults featuring ``strongly" locked ends and uniform shear stress loading, as would be the case for a fault of finite extent and examined here, or a fault patch which experiences non-uniform loading, such as loading by steady creep on one end and ``strong" or ``weak" locking at the other. The former, ``strong" locking may be due to a fault boundary beyond which the host rock is intact. In contrast, the latter, ``weak" locking may be due to a temporary, post-seismic seizing of fault regions that slipped seismically, locks with an effectively zero slip rate, but may eventually slide again.

\

Here we consider a fault undergoing in-plane or anti-plane deformation and ``strongly" locked at its ends $x=\pm L$ such that slip only occurs on $|x|<L$, while being uniformly loaded at a constant rate, which we denote as $\partial\tau_b/\partial t=\dot\tau_b$. In this case, we will consider perturbations about a steady-state condition, which for this case, will not have a uniform slip rate, but one that must satisfy (\ref{eq:tb}), or
\begin{equation}
\dot\tau_b + \frac{\mu'}{\pi}\int_{-L}^L\frac{V_o'(\xi)}{\xi-x}d\xi=0
\end{equation}
The solution to which is 
\begin{equation}
V_o(x)=\frac{\dot\tau_b L}{\mu'}\sqrt{1-(x/L)^2}
\label{eq:Vo2}
\end{equation}
where we identify for this problem the characteristic slip rate $V_{pl}$ as
\begin{equation}
V_{pl}=\frac{\dot\tau_b L}{\mu'}
\label{eq:Vpltdot}
\end{equation}
and the corresponding distribution of the state variable for steady-state sliding is $\theta_o(x)=D_c/V_o(x)$. 

\

We non-dimensionalize (\ref{eq:lin1}) and (\ref{eq:lin2}) in the manner of (\ref{eq:nd}), and rearrange to arrive to the pair of coupled, linear evolution equations for $\tilde V$ and $\tilde\theta$
\begin{subequations}
\begin{align}
\frac{\partial \tilde V}{\partial t}&= \frac{b}{a}\left(V_o(x)\mathcal{L}[\tilde V(x,t)] +V_o(x)\tilde V(x,t)+V^3_o(x)\tilde\theta(x,t)
\right) \\[9 pt]
\frac{\partial\tilde\theta}{\partial t}&=-V_o(x)\tilde\theta(x,t)-\frac{\tilde V(x,t)}{V_o(x)}
\end{align}
\label{eq:fsys1}
\end{subequations}
We again look for perturbations of the form (\ref{eq:pert}), non-dimensionalizing distance as $x/L\Rightarrow x$, introducing (\ref{eq:fin}), such that (\ref{eq:fsys1}) becomes
\begin{subequations}
\begin{align}
s \mathcal{V}(x)&= \frac{b}{a}\left[ V_o(x)\left(\frac{L_b}{2L}\frac{1}{\pi}\int_{-1}^1\frac{\mathcal{V}'(\xi)}{\xi-x}d\xi +\mathcal{V}(x)\right)+V^3_o(x)\Theta(x)
\right] \\[9 pt]
s\Theta(x)&=-\frac{\mathcal{V}(x)}{V_o(x)}-V_o(x)\Theta(x)
\end{align}
\label{eq:fsys2}
\end{subequations}
where we also require that $\mathcal{V}(\pm 1)=0$ to meet the locked boundary condition. We recognize (\ref{eq:fsys2}) as an eigenvalue problem with eigenvalue $s$, eigenfunctions $\mathcal{V}(x)$ and $\Theta(x)$, and parameters $a/b$ and $L/L_b$. Unlike (\ref{eq:fsys0}), we see that eliminating $\Theta(x)$ in (\ref{eq:fsys2}) does not simplify the problem and we instead solve the system (\ref{eq:fsys2}) numerically, using a procedure detailed in Appendix \ref{app:fin}. 

\begin{figure}[t]
      \center\includegraphics[width=400pt]{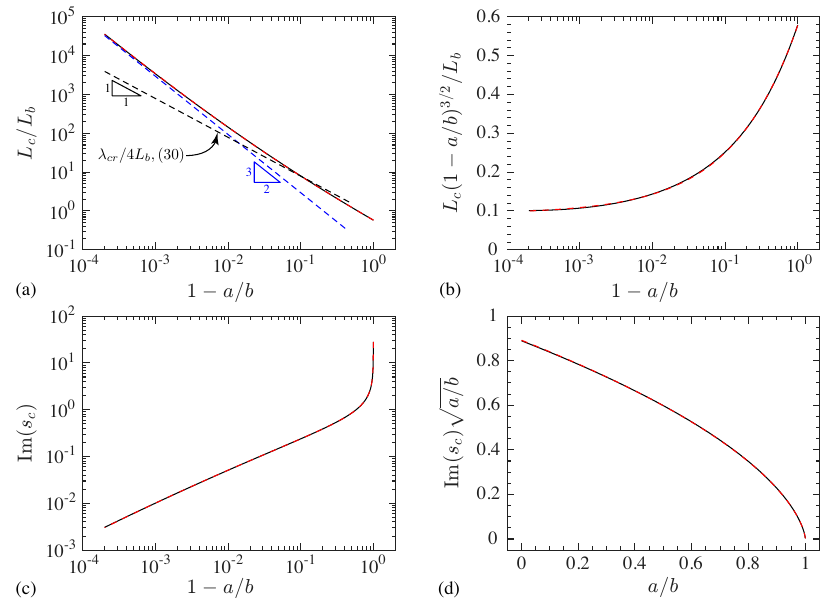}\vspace{5 mm}
   \caption{\raggedright Examination of the dependence of critical fault half-length and complex exponent on the problem parameter $a/b$ for a finite fault, locked at its edges and loaded by a constant rate of shear stress. (a,b) Plot of the critical fault half-length $L_c$ (solid black) and its approximation (red-dashed). (c,d) Plot of the imaginary part of the eigenvalue $s_c$ at stability loss (solid black) as well as its approximation (red dashed). For comparison in (a), we show the leading-order asymptotic behavior of the red-dash approximation in the limit  $a/b\rightarrow 1$ (blue-dashed), as well as the result of {\it Rice and Ruina} [1983] for the critical perturbation wavelength (black-dashed) for an infinite, homogeneous, rate-weakening fault, (\ref{eq:lamc}), reduced by a factor of four to compare the critical fault half-length with half the inter-nodal distance of an oscillation with wavelength $\lambda_c$.}
   \label{fig:finlin}
\end{figure}

\

We find that stability is lost at a critical value of the fault half-length, $L_c/L_b$, the value of which is a function of $a/b$. We find this critical half-length by fixing the value of $a/b$, increasing the parameter $L/L_b$ from 0, and solving for the spectrum of eigenvalues for each value of $L/L_b$. We observe that at the critical value, stability is lost as a pair of complex conjugate eigenvalues $s_c$ cross the imaginary axis. For $L<L_c$, the fault is linearly stable to perturbations. As $L$ is increased beyond $L_c$ a sequence of Hopf bifurcations occur at discrete values of $L$: additional unstable modes are present for longer faults. In Figure \ref{fig:finlin}a,b we show the numerical solution for the dependence of $L_c/L_b$ on $a/b$, found by a bisection method, as well as the approximation 

\begin{equation}
\frac{L_c}{L_b}\approx\frac{1}{(1-a/b)^{3/2}}\left[0.0929+0.503(1-a/b)^{1/2}-0.0178(1-a/b)\right]
\label{eq:Lc2}
\end{equation}

\noindent which has $<1$\% relative error for $0<a/b<0.998$. While this has the appearance of an asymptotic expansion in the limit $a/b\rightarrow 1$, we have not conducted an asymptotic analysis of the eigenvalue problem at the Hopf-bifurcation point. However, the numerical solutions do suggest the leading-order asymptotic term $O((1-a/b)^{-3/2})$ in the limit $1-a/b\rightarrow 0$. The remaining terms in (\ref{eq:Lc2}) are chosen to provide an accurate, low-parameter fit of the numerical results over the entire range of $a/b$ examined numerically.

\

We emphasize here that the apparent asymptotic scaling of $L_c/L_b$ changes when shifting from a uniform to non-uniform baseline rate of sliding: the critical fault dimension $L_c/L_b$ is $O((1-a/b)^{-3/2})$ as $a/b\rightarrow 1$ here, representing an abrupt departure from the $(1-a/b)^{-1}$ scaling seen in section \ref{sec:uni} for the uniformly sliding infinite fault, (\ref{eq:lamc}), and the uniformly sliding asperity, (\ref{eq:Lc1}).

\

In Figure \ref{fig:finlin}c,d we show the value of imaginary part of the eigenvalue $s$ at the Hopf bifurcation point, as well as its approximation
\begin{equation}
\text{Im}(s_c)\approx\pm0.892\sqrt{\frac{b}{a}}(1-a/b)^{0.569} e^{-0.0115[\ln(1-a/b)]^2}
\label{eq:sc2}
\end{equation}
which has $<2$\% relative error for $0<a/b<0.1$ and $<1$\% relative error for $0.1<a/b<0.999$. The approximation (\ref{eq:sc2}) makes an explicit attempt at matching an apparent leading-order $\sqrt{b/a}$ asymptotic behavior of $\text{Im}(s_c)$ as $a/b\rightarrow 0$, an asymptotic scaling also seen in (\ref{eq:sc1}). However, the following terms in (\ref{eq:sc2}) are chosen to optimize fit over the entire range of $a/b$ explored in numerical solutions and are not to be taken to follow from an asymptotic analysis.

\section{Nucleation lengths from a non-linear analysis of instability}
\label{sec:nonlin}
There is a temptation to use the critical lengths $L_c$ calculated in sections \ref{sec:uni} and \ref{sec:finuni} as a nucleation length, but this is a fine point: one should be clear as to what one is considering the nucleation of. Here we define the nucleation length as the initial size, or nucleus, of a dynamic, earthquake-generating rupture. By this definition, $L_c$ cannot be qualified as a nucleation length. The linear stability analysis only indicates whether spontaneous acceleration beyond a specific state of steady-state sliding can initiate provided the fault dimensions are above a critical size. This initially unstable acceleration may or may not progress to seismic slip speeds. The stability analysis assumes slip rate and state are within a linear regime, in which perturbations from steady-state are only within a few percent of steady-state values and a linearized frictional response is used. However, seismic slip speeds of O(m/s) are ten orders of magnitude above background plate rates of O(cm/yr). Thus in order to consider the earliest phase of a dynamic regime, one must be able to examine the quasi-static acceleration of the fault well into a non-linear regime, which then becomes dependent on the specific non-linear state evolution law assumed.

\subsection{``Free-boundary" aging-law nucleation}
\label{sec:free}
A series of works [{\it Rubin and Ampuero}, 2005; {\it Ampuero and Rubin}, 2008; {\it Rubin and Ampuero}, 2009; {\it Viesca}, 2016a,b] has examined this non-linear phase of acceleration to determine nucleation lengths. {\it Viesca} [2016a,b, 2023] demonstrated that an analysis commonly used for non-linear, time-dependent problems can be applied to unstable acceleration of faults with rate-and-state friction with evolution laws ranging from aging to slip. Specifically, {\it Rubin and Ampuero} [2005] and {\it Viesca} [2016a,b, 2023] sought solutions in which slip rate diverges within a finite time $t_o$ in the separable form 
\begin{equation}
V(x,t)=\frac{D_c}{t_o-t}\mathcal{W}(x)
\label{eq:bu}
\end{equation}
in which the spatial distribution $\mathcal{W}(x)$ is solved for and found to be a compact distribution within $|x|\leq L_f$.
For in-plane or anti-plane sliding between two elastic half-spaces on a fault with uniform friction properties in the rate-weakening range $0<a/b<1$, {\it Viesca} [2016a,b] showed that the spatial distribution of diverging slip $\mathcal{W}$ and the distance $L_f$ must satisfy
\begin{equation}
1-\frac{a}{b}+\frac{L_b}{L_f}\frac{1}{2\pi}\int_{-1}^1 \frac{\mathcal{W}'(s)}{s-x}ds=\begin{cases}1-\mathcal{W}(x) & \mathcal{W}\leq 1 \\ 0 & \mathcal{W}\geq 1\end{cases} 
\label{eq:W}
\end{equation}
where we have nondimensionalized distance as $x/L_f\Rightarrow x$. This problem is a free-boundary problem: the size of the region over which slip rate diverges is given by $L_f$, an unknown to be solved for. Equation (\ref{eq:W}) is not sufficient to determine $\mathcal{W}(x)$ and $L_f$: an additional constraint is required. This constraint is the requirement that there be no stress-rate singularity just ahead of the tips of the compact region at $x=\pm L_f$. The physical origin of this requirement is the presumption that unstable acceleration of slip is occurring on a region contained within a larger fault whose finite strength cannot support such a singularity. This requirement amounted to imposing the condition that $\mathcal{W}'(\pm 1)=0$. {\it Viesca} [2016a,b] solved (\ref{eq:W}) for  $\mathcal{W}$ and the dimensionless size $L_f/L_b$, as they both depend on the parameter $a/b$. The dependence of $L_f/L_b$ on $a/b$ is plotted as black curves in Figure \ref{fig:nonlin}. The solution for $L_f$ may be approximated  by the following expression
\begin{equation}
\frac{L_f}{L_b}\approx\begin{cases}
1.377377286... &0<a/b<0.3780720054...\\ 
\frac{1}{\pi(1-a/b)^2}\left[1+1.11(1-a/b)^2-0.541(1-a/b)^4+5.53(1-a/b)^6\right] & 0.3781...<a/b<1
\end{cases}
\label{eq:Lf}
\end{equation}
where the first case is precise to as many significant digits shown and the latter case has a relative error less than 1\%, such that if the approximation were overlaid on the solutions for $L_f$ in Figure \ref{fig:nonlin}, it would be indistinguishable.
Remarkably, these solutions retrieved two lengthscales proposed earlier by {\it Rubin and Ampuero} [2005]
\begin{align}
L_f=(1.3773...)L_b \quad &a/b<0.3781...\\[9 pt]
L_f\sim\frac{L_b}{\pi(1-a/b)^2} \quad &a/b\rightarrow 1
\end{align}
The first lengthscale was found by {\it Rubin and Ampuero} [2005] and appears as a special case of the general analysis pursued by {\it Viesca} [2016a,b] and the second by the former authors approximating instability development via a fracture analogy. {\it Viesca} [2016a,b] provided a consistent analysis of nucleation over the entire range of $a/b$ and showed that the second lengthscale is, in fact, the appropriate asymptotic limit for $L_f$ as $a/b\rightarrow 1$, deriving the equivalent fracture problem $L_f$ must satisfy. $L_f$ can be considered as the expected nucleation length on an aging-law fault provided the nucleation patch of size $2L_f$ is smaller than the fault or its frictionally weakening asperity potentially capable of hosting sliding instabilities.

\subsection{``Pinned" aging-law nucleation}
\label{sec:pin}
\begin{figure}
      \center\includegraphics[width=400pt]{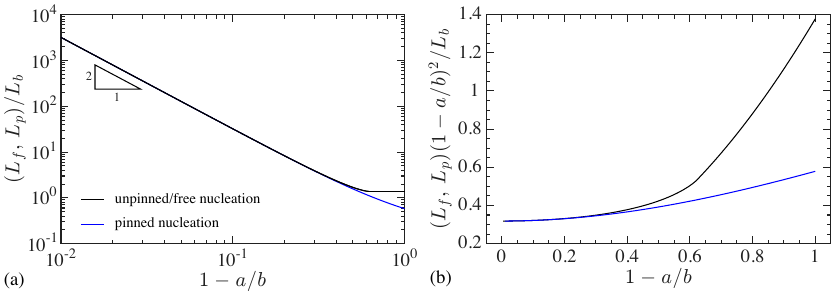}\vspace{5 mm}
   \caption{\raggedright Illustration of the dependence on $a/b$ of free and pinned nucleation lengths, under in-plane or anti-plane conditions, and their convergence as $a/b\rightarrow 1$. (a,b) Plot of the free $L_f$ (solid black) and pinned $L_p$ (solid blue) nucleation lengths for an aging-law fault, determined by solution for the compact support of the function $\mathcal{W}(x)$ in the non-linear instability solutions (\ref{eq:bu}), under singularity-free (free) and singularity-present (pinned) conditions. Approximations of the free and pinned nucleation lengths, (\ref{eq:Lf}) and (\ref{eq:Lp}), though not plotted, are indistinguishable from the solid curves here.}

   \label{fig:nonlin}
\end{figure}

The finiteness of a fault or its frictionally weakening asperity may play a role in determining the nucleation length. The results of section \ref{sec:free} presumed that a singularity in stress rate was inadmissible and that the rate-weakening portion of the fault (asperity) extended beyond the nucleation zone. However, for sufficiently small faults, nucleation and dynamic rupture may be contained within the fault if the surroundings are able to support a singularity at fault boundaries: such cases may occur for a finite-sized rate-weakening fault that is locked at its edges, or for a finite rate-weakening asperity surrounded by steady creep. In the former case, the fault terminates and is surrounded by rock with finite toughness capable of sustaining a singularity. In the latter, the rate-weakening asperity lies within a steadily creeping region and that surrounding region may be strongly rate-strengthening (i.e., in the limit $a/b\rightarrow \infty$) or the fault exterior to the asperity is simply restricted to slide at a constant rate by model design. In any case, we may search for so-called ``pinned" non-linear solutions of the form (\ref{eq:bu}) again solving (\ref{eq:W}) (with $L_f$ therein replaced with $L_{p}$ and $x$ implicitly scaled by $L_{p}$) and removing the requirement imposed by {\it Viesca} [2016a,b] that $\mathcal{W}'(\pm 1)=0$. Here $L_{p}/L_b$ is no longer a solution variable determined by a boundary condition but now can be considered as a free parameter. Restricting ourselves to solutions for which $\mathcal{W}$ (and hence $V$) is positive within a region $|x|\leq 1$, we find that the pinned solutions occupy a range $L^{min}_{p}<L_{p}<L_f$, where $L^{min}_p$ is the lowest possible value for which a solution is found to exist (see Appendix \ref{app:pin}). Since this lower bound is the most relevant value in future discussions, we will drop the ``min" superscript and simply dub this minimum ``pinned" nucleation length $L_p$, letting it be understood that there exist a family of pinned solutions starting from this minimum value and up to $L_f$. The dependence of $L_p/L_b$ on $a/b$ is plotted as blue curves in Figure \ref{fig:nonlin}.  $L_{p}$ may be approximated, to less than 0.1\% relative error over the rate-weakening range $0<a/b<0.999$, by
\begin{equation}\frac{L_{p}}{L_b}\approx \frac{1}{\pi(1-a/b)^2}\left[1+0.974(1-a/b)^2-0.199(1-a/b)^4+0.0440(1-a/b)^6\right]\label{eq:Lp}\end{equation}
The accuracy of the approximation is sufficient to make it indistinguishable from the blue-curves in Figure \ref{fig:nonlin}.
When the direct effect is neglected ($a/b=0$), (\ref{eq:W}) reduces to the Uenishi-Rice eigenvalue problem in which the physically admissible solution is that of the lowest eigenmode: i.e., $L_p/L_b=0.57888694...$ in that case.

$2L_p$ represents the size of the smallest asperity within a creeping region or the smallest fault that can sustain a non-linear instability towards seismic slip speeds, if the fault has uniform rate-weakening properties. This may be considered a nucleation length in that it marks the starting size for a dynamic rupture. However, under the conditions described above---locked edges, or rate-strengthening or steadily creeping surroundings---the dynamic rupture would be expected to be contained within the nucleation area, unless another weakening mechanism is enabled by the high slip and stress rates of the dynamic rupture to allow for dynamic rupture to penetrate the surrounding barriers.

\subsection{``Free" and ``pinned" nucleation from aging to slip laws}
\label{sec:se}
The nucleation lengths for unpinned (\ref{eq:Lf}) and pinned (\ref{eq:Lp}) faults were found assuming that the fault followed the aging state evolution law (\ref{eq:ag}), but the analysis and results can be extended to friction laws ranging between aging and slip laws. Using the intermediate law (\ref{eq:in}), whose parameter $\epsilon$ allows a continuous transition between aging ($\epsilon=1$) and slip ($\epsilon\rightarrow 0$) laws, {\it Viesca} [2023]  showed that non-linear blow-up solutions of the form (\ref{eq:bu}) continue to exist for the intermediate law. This is seen by repeating the derivation of {\it Viesca} [2016a,b] leading to (\ref{eq:W}), except introducing a factor of $\epsilon$ to the right hand side of (\ref{eq:bu}).  Doing so, {\it Viesca} [2023] found that the equation governing $\mathcal{W}$ is exactly that for the aging law, (\ref{eq:W}), except the lengthscale $L_f$ (or $L_p$) is reduced by a factor $\epsilon$. This means that the expressions for nucleation lengths $L_f$ and $L_p$ also apply for state evolution laws ranging from aging to slip, provided the right hand sides of (\ref{eq:Lf}) and (\ref{eq:Lp}) are multiplied by $\epsilon$. 

\

An immediately apparent consequence of this result is that nucleation lengths vanish as $\epsilon\rightarrow 0$. This is borne out in numerical solutions of quasi-static instability on slip-law faults [{\it Ampuero and Rubin}, 2008; {\it Rubin and Ampuero}, 2009], in which the slip rate simultaneously accelerates and contracts towards a point in space in finite time. Those results also show apparent migration of the accelerating slip in the form of a slip-pulse. The slip-pulse nature was shown by {\it Viesca} [2023] to emerge as the consequence of the partial loss of stability of the blow-up solution (\ref{eq:bu}) for rate-weakening values of $a/b>0.3781...$, such that migration occurs, but the point-like nature of the blow-up is preserved. In fully dynamic models of the rupture process, quasi-static acceleration is ultimately limited by inertia, which prohibits continued spatial collapse of acceleration and induces outward propagation of a dynamic rupture.

\section{Transition of fault slip acceleration from linear to non-linear stages}
\label{sec:trans}

Linear stability analysis is initial-conditions-specific. The initial conditions in the form of the underlying steady-state distribution of slip rate and state, about which perturbations are considered, are a reflection of the fault geometry (e.g., finiteness), possible frictional heterogeneity, and boundary and loading conditions.  As shown in section \ref{sec:linstab}, the critical length $L_c$ is strongly determined by the initial spatial distribution of slip rate and state and we found that the asymptotic scaling of $L_c/L_b$ in the limit $a/b\rightarrow 1$ changes with the form of the initial condition. That scaling changes from $O((1-a/b)^{-1})$ for a finite rate-weakening asperity driven by boundary creep to $O((1-a/b)^{-3/2})$ for uniform rate of shear stress loading on a finite fault. The only commonality of the stability results for the two loading configurations occurs when the direct effect is neglected ($a/b=0$), in which case the critical length $L_c/L_b$ for both configurations is given by the first {\it Uenishi and Rice} [2003] eigenvalue ($0.57888694...$). An additional signature of the initial and loading conditions is the characteristic slip velocity ($V_{pl}$) and timescale ($D_c/V_{pl}$) for evolution of the slip rate following perturbation. For example $V_{pl}$ in the case of constant rate of shear stress, is given by (\ref{eq:Vpltdot}), whereas for the case of loading by steady creep, $V_{pl}$ reflects the rate of slip in the boundary regions.

In contrast, in a non-linear regime, finite-time divergence of slip rate in the form (\ref{eq:bu}) is independent of fault slip history, and, by extension, of particular loading and fault initial conditions of slip rate and state. The characteristic timescale and slip rate contain no information on the process that triggered instability. Rather, the relevant timescale is given by the time leading up to a finite-time divergence $t_f(t)=t_o-t$ and the relevant slip rate scale is $D_c/t_f(t)$, which diverges as $t\rightarrow t_o$. The ``free" and ``pinned" nucleation lengths $L_f$ and $L_p$, as well as the slip rate distribution $\mathcal{W}$ also contain no information of initial conditions. This independence implies that late-stage acceleration of fault sliding exhibits a type of universality: there exists a pathway for slip rate to continue to accelerate unstably regardless of what triggered the instability. An exception to this universality occurs when solutions in the form (\ref{eq:bu}) are no longer asymptotically stable [{\it Viesca}, 2016a,b]. In such cases, non-linear divergence occurs, though with time-dependent deviations from divergence in the form of (\ref{eq:bu}) and initial conditions exert an influence on these deviations. It should also be noted that, while not influencing the existence of this universal pathway, the distribution of effective normal stress on the fault during this divergence does exert an influence on the distribution of slip rate during finite-time divergence, $\mathcal{W}$, as well as the nucleation lengths $L_f$, and $L_p$. This influence is discussed further by {\it Ray and Viesca} [2017].

\

How does an instability on the fault transition from an unstable acceleration determined by fault initial and loading conditions to the one independent of it? In other words, what ensures that this universal terminal pathway is taken? For example, consider the initial growth of perturbations from an initial steady-state: $V=V_o(x)+\tilde V(x,t)$ with $\tilde V(x,t)=\mathcal{V}(x)\exp(st)$. Let's follow the fastest growing mode $\mathcal{V}_1(x)$ with corresponding $s_1$. We recall that this mode shape and exponential growth rate presumed small departures such that the governing equation could be reasonably linearized. We also recall that the form of $V_o(x)$ and $\mathcal{V}_1(x)$ and the exponential growth rate $s_1$ are determined by the manner of initial and loading conditions, the fault size $L/L_b$, and ratio $a/b$. For linear instability, the first two parameters are such that $\text{Re}(s_1)>0$. However, as acceleration continues, non-linear effects become important and the distribution of slip rate and acceleration will depart from this modal exponential growth. 

Why might slip rate transition to the diverging nucleation solution $V(x,t)=\mathcal{W}(x) D_c/(t_o-t)$? That slip rate would favor approaching this form follows from a dynamical-systems analysis of its asymptotic stability [{\it Viesca}, 2016a,b]. This analysis indicated that under most cases, solutions of the form $V(x,t)=\mathcal{W}(x) D_c/(t_o-t)$ are attractive and asymptotically stable, meaning that they are expected to be approached in the limit $t\rightarrow t_o$. This asymptotic stability is readily observable under the aging-law for $a/b<0.74$. In limited cases, for which $a/b$ is in the vicinity of unity, these solutions are not asymptotically stable, but may be considered to be Lyapunov stable and attractive nonetheless; this is most noticeable in numerical solutions for aging-law with $a/b>0.9$, in which the presence of several unstable modes leads to alternation between slip rate approaching the solution (\ref{eq:bu}) and large, chaotic deviations. This analysis reconciled the apparent dichotomy in behavior of earlier numerical solutions that were observed to exhibit asymptotic stability for $a/b<0.3781...$ and a lack of it for $a/b$ near 1 [{\it Rubin and Ampuero}, 2005]. 

\begin{figure}[t]
      \center\includegraphics[width=200pt]{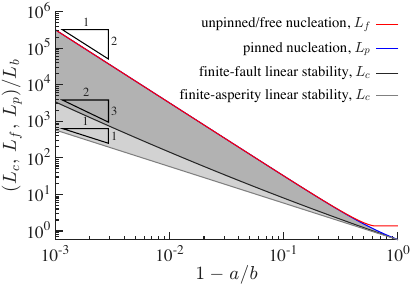}\vspace{5 mm}
   \caption{\raggedright Comparison of linear-stability thresholds for the two finite-fault loading conditions and the non-linear nucleation lengths. Plot of the critical fault half-length $L_c$ at which point linear stability is lost for two steady-state initial conditions: steady creep at the boundary of an asperity (dark grey), finite-fault loaded at a constant rate of shear stress (black), compared with the nucleation lengths $L_f$ (red) and $L_p$ (blue). For both loading cases $L_c<L_p<L_f$ for all $0<a/b<1$. Shaded regions highlight areas bounded by $L_c$ and $L_p$.}
   \label{fig:Lcomp}
\end{figure}

\

However, the following question remains to be answered: under what conditions will a linear instability progress beyond its linear phase and accelerate within a non-linear regime until a dynamic rupture is nucleated? Posed alternatively, does a system that is linearly unstable necessarily continue to accelerate? This question might be most easily discussed considering fault length $L$ as a variable, with all other parameters held fixed. In this case, we can simply compare $L$ to the critical linear stability length $L_c$ and ``free" or pinned nucleation lengths $L_f$ and $L_p$. 

\

As a starting point, we compare $L_c$ among the two cases of linear stability involving a finite fault dimension $L$: (i) a fault frictional asperity of length $2L$ uniformly creeping at steady-state and loaded by a constant slip rate at the boundaries; (ii) a finite fault of length $2L$ loaded with a constant rate of shear stress. Using here a superscript to denote the $L_c$ corresponding to each case, the relative values of $L_c$ follow a clear order, $L^{(i)}_c<L^{(ii)}_c$, for all $0<a/b<1$ (Figure \ref{fig:Lcomp}). 

\

\begin{figure}
      \center\includegraphics[width=400pt]{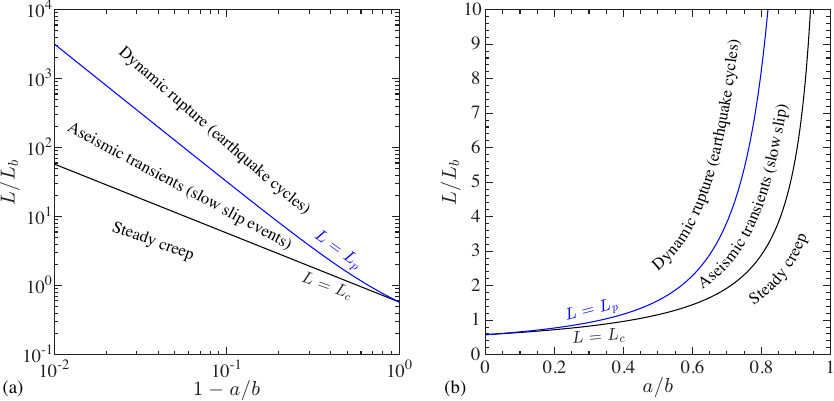}\vspace{5 mm}
   \caption{\raggedright Phase diagrams for the expected response of an aging-law rate-weakening fault within $|x|<L$, driven by a constant rate of creep $V_{pl}$ along $|x|>L$. The response may be either steady, uniform creep at $V=V_{pl}$ for $L<L_c$, aseismic transients for $L_c<L<L_p$, or the emergence of dynamic rupture for $L_p<L$. The phase boundaries are determined by the linear stability critical fault size $L_c$, (\ref{eq:Lc1}), and minimum pinned nucleation length $L_p$, (\ref{eq:Lp}). While $L_c$ is independent of the chosen state evolution law (aging, slip or an intermediate law), $L_p$ is dependent on that choice; here $L_p$ is drawn for aging-law state evolution.}
   \label{fig:phase}
\end{figure}

We now compare $L_c$ for the two cases with two nucleation lengths, $L_p$ and $L_f$, which have their own clear order, $L_p<L_f$. In Figure \ref{fig:Lcomp} we show that $L_c^{(i)}<L_c^{(ii)}< L_p<L_f$ for all $0<a/b<1$. We now return to the problem of considering a variable $L$ with all other parameters fixed. For one of the two $L_c$, if $L<L_c$, the fault is linearly stable to any small perturbations to the initial steady state. It is then also necessarily the case that $L< L_p$, the smallest possible nucleation length. Thus at this fault length $L$, there exists no solution for finite-time divergence of slip rate accounting for the full nonlinearity of the evolution equations. This implies that the fault is also nonlinearly stable: there is no finite (large) perturbation that will provoke the system to unstably accelerate. In other words, for $L<L_c$ the fault can be expected to remain at the steady-state sliding corresponding to its loading. There will be no spontaneous, transient acceleration, other than that directly imposed by an applied perturbation, and such an acceleration will ultimately decay, with fault slip returning to the initial steady sliding. For fault lengths $L$ in the window $L_c<L<L_p$, the fault can be considered to be linearly unstable but non-linearly stable. This means that small perturbations to steady state can grow with temporary excursions into the non-linear regime. However, because the fault is not sufficiently large to admit solutions that diverge in this non-linear regime, these excursions cannot transition into spontaneous acceleration and diverging slip rate. A likely outcome is that the fault will ultimately decelerate back towards steady-state, before another perturbation repeats this process. The clear delimitation of critical linear and non-linear fault lengths and phase boundaries seen in  Figures \ref{fig:Lcomp} and \ref{fig:phase} occurs for the aging law. As mentioned in section \ref{sec:se}, the free and pinned nucleation lengths are reduced by a factor $\epsilon$ for a state evolution law that transitions from aging to slip. This leads to a cross-over of the linear and non-linear lengths. We defer a discussion of this point and its implications to section \ref{sec:discse}.

\section{Discussion}
\label{sec:disc}

\subsection{Implications for seismic cycle models}

How do the results of sections \ref{sec:linstab}--\ref{sec:trans} inform models of the seismic cycle coupling elastic deformation of the host rock with rate- and state-dependent friction on the fault? Here we focus our discussion on the specific case of a rate-weakening fault patch embedded within an otherwise stably creeping fault, though the discussion would be similar for the other case for which we considered linear stability. 

\

In Figure \ref{fig:phase} we draw a phase diagram of the response of the fault patch for a particular scaled size of the patch, $L/L_b$, and relative degree of rate-weakening $a/b$. The three possible behaviors of the fault patch delimited by the phase boundaries are either steady creep, the spontaneous emergence of large-amplitude aseismic transients without dynamic rupture, and the spontaneous nucleation of dynamic rupture. The boundaries delimiting the three responses are determined by the linear stability length, $L_c$, and the minimum nucleation length, $L_p$, given as functions of $a/b$ by equations (\ref{eq:Lc1}) and (\ref{eq:Lp}), respectively. The expression of $L_p$ is assuming aging-law state evolution; a prefactor of $\epsilon$ must be introduced if assuming the intermediate evolution law (\ref{eq:in}). When $L<L_c$, steady sliding of the rate-weakening patch at a uniform rate $V_{pl}$ is both linearly and non-linearly stable: any perturbations to the fault will ultimately decay and the fault will return to steady creep.

\

For $L$ in the range $L_c<L<L_p$ we may expect the spontaneous emergence of aseismic transients: the fault is linearly unstable meaning small perturbations will grow exponentially quickly but this result of linearization loses relevance as the growth of slip rate following a small perturbation enters the non-linear regime. Exponential growth of perturbations does not continue indefinitely, and provided $L$ remains below $L_p$, there does not exist an attractive manner for slip to continue to accelerate. This means that any transient, finite acceleration induced by being linearly unstable is ultimately limited as the fault is non-linearly stable, and the fault will decelerate towards steady-state and the process repeats. 

\

When $L>L_p$, the fault can spontaneously accelerate towards dynamic rupture. In this case, the fault is linearly unstable, small perturbations spontaneously grow. Because the fault is now also nonlinearly unstable, those perturbations may continue to grow in the non-linear regime. In other words, the fault is sufficiently large to admit attractive pathway for the slip rate to diverge in this non-linear regime, in the form of the solution (\ref{eq:bu}). For $L$ in the range $L_p\leq L\leq L_f$, that acceleration towards inertial sliding and dynamic rupture occurs over the entire fault length $L$. This means that sliding reaches seismic slip velocities simultaneously over the entire fault and there is no outward propagation of this dynamic rupture, since in this model, sliding beyond $x=\pm L$ occurs at a fixed rate $V_{pl}$; the rate-weakening patch will slip seismically, lock, and will subsequently be reloaded by the boundary creep to repeat the process. For $L$ above $L_f$ (not shown on the phase diagram of Figure \ref{fig:phase}, but asymptotically approaching $L_p$ in the limit $a/b\rightarrow 1$, as shown in Figure \ref{fig:nonlin}), the dynamic rupture can propagate before arresting at the boundaries $x=\pm L$. 

\

Both the linear stability and non-linear nucleation analyses here can be adapted to account for heterogeneity of frictional properties and effective normal stress. Seismic cycle models typically include variations in the coefficients $a$ and $b$ and also normal stress. Variations of the former typically take the shape of a transition from rate-weakening to rate-strengthening behavior down dip. Such variations were originally in early models of the seismic cycle [{\it Tse and Rice}, 1986; {\it Rice}, 1993; {\it Lapusta et al.}, 2000], justified on the basis of hydrothermal dependence of frictional properties [{\it Stesky}, 1975; {\it Blanpied et al.}, 1991, 1995]. Smooth or abrupt variations of frictional properties are now regularly presumed on model rate-and-state faults and work has begun to identify stability transitions due to specific forms of heterogeneity. For example, prior work has considered faults  with alternating variations of rate-strengthening and rate-weakening properties [e.g., {\it Yabe and Ide}, 2017] in elastic or visco-elastic continua and performed analyses of stability transitions via approximations of the continuum models with interacting spring-block models (with damping in the viscoelastic case) that prove faithful to transitions observed in the corresponding continuum models [{\it Luo and Ampuero}, 2018; {\it Noda and Yamamoto}, 2024]. For arbitrary variations of properties, such a reduction may not always be possible, and for a fault in a continuum, one may repeat the linear stability analyses of section \ref{sec:linstab} accounting for the heterogeneity of $a$ and $b$, solving for the eigenfunctions $\mathcal{V}$ and $\Theta$ and eigenvalues $s$ numerically. We defer this to subsequent work. Regarding unstable acceleration in the non-linear regime, {\it Ray and Viesca} [2017] considered the problem of finding solutions of the form (\ref{eq:bu}) and the nucleation length $L_f$ considering heterogeneity of frictional properties $a$, $b$, and $D_c$. They demonstrated that the equation governing the slip-rate distribution $\mathcal{W}$ and $L_f$ is equivalent to a classical fracture problem of a slip-weakening crack in equilibrium with an applied load: thus, results and techniques developed considering heterogeneity in slip-weakening friction have extensions to determine nucleation lengths under heterogeneity of rate-and-state frictional properties [e.g., {\it Lebihain et al.}, 2021]. Subsequent work [{\it Ray and Viesca}, 2019], demonstrated the possibility of homogenization in which property variations occur over wavelengths disparate from elasto-frictional lengthscales (such as $L_b$ appearing in this work). While {\it Ray and Viesca} [2017, 2019] considered the so-called ``free-boundary" nucleation ($L_f$), the governing equations remain the same for pinned nucleation $L_p$ lengths under heterogeneity. Solving for $L_p$ under heterogeneity can be done as pursued in section \ref{sec:pin} and Appendix \ref{app:pin}: by removing the condition that $\mathcal{W}'$ vanish at its endpoints. 

\begin{figure}
      \center\includegraphics[width=400pt]{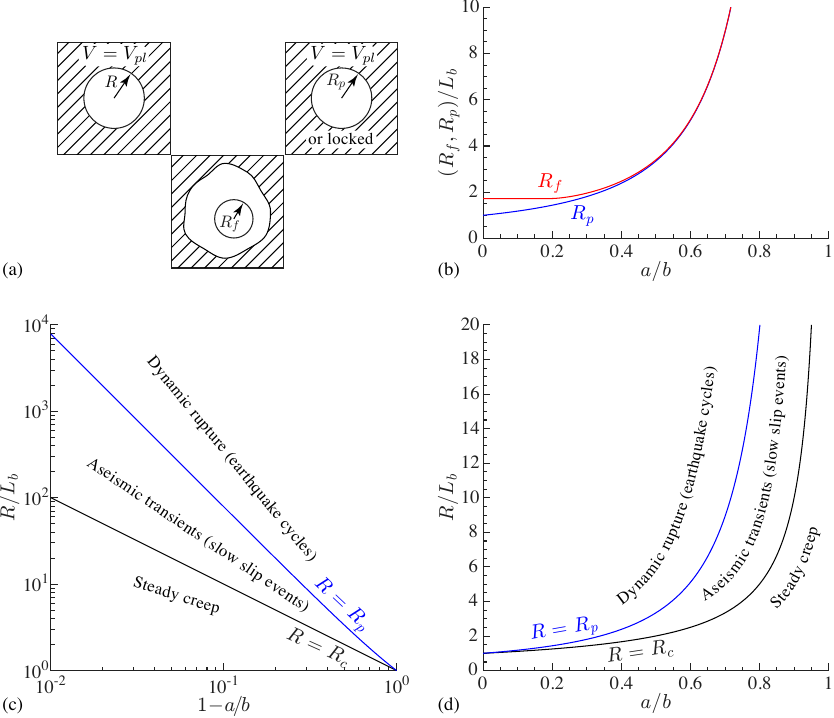}\vspace{5 mm}
   \caption{\raggedright Schematics and summary of results concerning planar faults under conditions of deformation in three dimensions (mixed-mode sliding). (a) Schematic drawing of conditions (left) of a circular rate-weakening patch with a radius $R$ and within a fault that otherwise creeps at constant rate $V_{pl}$; (right) the smallest radius $R_p$ of a circular rate-weakening patch, confined within an otherwise creeping or locked fault, that may host a runaway, non-linear instability; and (bottom) a circular nucleating region with radius $R_f$ that can exist within a larger, rate-weakening fault. (b) Comparison of $R_f$ and $R_p$ as they depend on $a/b$: for any $0<a/b<1$, $R_p<R_f$, and the two asymptotically approach each other as $a/b\rightarrow 1$. (c,d) Phase diagram for expected behavior of fault models containing a circular rate-weakening region of radius $R$ within an otherwise steadily creeping fault. For $R<R_c$, the fault is linearly and non-linearly stable, and steady creep is an unconditionally stable solution. For $R_c<R<R_p$ the fault is linearly unstable, but non-linearly stable: the fault exhibits spontaneous emergence of aseismic transients, but acceleration is ultimately limited. For $R>R_p$ the fault is both linearly and non-linearly unstable and acceleration can spontaneously progress towards seismic slip speeds and the nucleation of dynamic rupture.}
   \label{fig:3d}
\end{figure}

\subsection{Linear stability and nucleation lengths in 3D (planar faults)}
\label{sec:3d}
Up to now, in sections \ref{sec:linstab}--\ref{sec:trans} we considered a single mode of slip, in-plane (mode-II) or anti-plane (mode-III), in which variations occurred along a single dimension and boundaries were represented by points. How do results change if a fault is undergoing mixed-mode slip, in which variations can occur in both dimensions of a two-dimensional fault plane and the boundaries are now curves? Here we are interested in determining critical dimensions under such conditions. The simplest geometry to consider under mixed-mode conditions is a circular region of radius $R$. We will present a solution for the critical radius, $R_c$, of a rate-weakening patch at which point its linear stability is lost. For brevity, we will do so for the specific loading case analogous to the single-mode problem considered in section \ref{sec:fin}. In addition we will present solutions for the free and pinned nucleation radii $R_f$ and $R_p$. The former corresponds to the size of a circular patch that can nucleate dynamic rupture within a larger, rate-weakening fault, which itself is not necessarily circular. The latter corresponds to the smallest circular rate-weakening region that may nucleate dynamic rupture while confined at its edges, because  the region exterior to the rate-weakening patch is either locked or undergoing an imposed slip rate. 

\

Variations of such a simple geometry have been the subject of prior numerical studies. Among the earliest works, {\it Kato} [2003] was explicitly concerned with the emergence of aseismic and seismic transients in a circular rate-weakening asperity embedded within a rate-strengthening fault and used circular crack solutions to provide a first estimate of a relevant radial lengthscale. A similar model geometry was subsequently used by {\it Veedu et al.} [2020] with finer parameter variations to document sharp transitions from stable sliding to aseismic transients and again to seismic sliding in parameter space. {\it Chen and Lapusta} [2009], followed by {\it Cattania and Segall} [2019] and {\it Chen and Lapusta} [2019], also considered a circular asperity model as a means to explain the peculiar moment-duration scaling of observed repeating earthquakes, though simulations were intentionally limited to parameter space that resulted in seismic events. In these repeating-earthquake works, those authors used fracture-mechanics based extension of results of {\it Rubin and Ampuero} [2005] to estimate a critical boundary for seismic events. For simplicity, we will neglect the presence of an outer rate-strengthening region, though our analysis may be easily extended to include it in later work.

\

For a circular region with an axisymmetric distribution of uni-directional slip within $r<R$, the traction on the fault plane is only parallel to that slip provided the Poisson ratio $\nu=0$. In this case, the operator $\mathcal L$ in (\ref{eq:t}) is given by [{\it Salamon and Dundurs}, 1971, 1977; {\it Bhattacharya and Viesca}, 2019]
\begin{equation}
\mathcal{L}[\delta(r,t)]=\frac{\mu'}{\pi}\int_0^R \left[\frac{E[k(r/s)]}{s-r}+\frac{K[k(r/s)]}{s+r}\right]\frac{\partial \delta(s,t)}{\partial s}ds
\label{eq:Lr}
\end{equation}
where $\mu'=\mu/2$, $E$ and $K$ are the complete elliptic integrals of the second and first kind, and $k(u)=2\sqrt{u}/(1+u)$ is the elliptic modulus. While $\nu=0$ may appear to be a peculiar case to be concerned with, it can be considered in terms of a perturbation series solution to a problem in which $\nu\neq 0$ and for which $\nu$ is the perturbation parameter. In this context, the $\nu=0$ solution is the leading-order approximation to the $\nu\neq0$ solution. For shear cracks, this was first exhibited for the problem of a singular crack in equilibrium with a uniform loading and having a uniform energy release rate (matching a uniform fracture energy) along its boundary, analytically treated by {\it Gao} [1988], following {\it Rice} [1985]. There the leading-order ($\nu=0$) solution to this mixed-mode problem is a circular crack. The next-order correction to the boundary shape and slip distribution is of the order $\nu$. This is also seen for non-singular ruptures. An example of a problem in this vein is a fault with constant Coulomb friction, the rupture of which is driven by the migration of fluid along the fault away from a point source injecting at constant rate [{\it S\'aez et al.} 2022; {\it Viesca}, 2025]. Closed-form expressions for the leading-order ($\nu=0$) circular-crack solution to this problem were provided as a perturbation expansion by {\it Viesca} [2025] and comparisons of these with numerical solutions to the problem in which $\nu>0$ [{\it S\'aez et al.}, 2022, 2025] show that the perturbations to the circular shape and peak slip are, to leading order, linear in $\nu$. Remarkably while some terms require correction, the authors found, to within the accuracy of the numerical solutions, that the evolution of the fracture area is preserved and independent of $\nu$. Thus the solution for which $\nu=0$ is an informative and often adequate basis to consider the three-dimensional effects on rupture.

\

We now look to extend linear stability results for a single mode of slip to the mixed-mode problem, taking as an example the mixed-mode equivalent to the problem of section \ref{sec:fin}: a circular rate-weakening fault patch or asperity that has radius $R$ and is surrounded by steady creep at the rate $V_{pl}$ (Figure \ref{fig:3d}a, left). The steady-state solution is uniform, steady sliding at the rate $V_{pl}$ within $r<R$. As part of the class of problems of section \ref{sec:uni}, the critical radius $R_c$ is determined by first solving the eigenvalue problem (\ref{eq:eig}) with $\mathcal{L}$ therein as given by (\ref{eq:Lr}). Non-dimensionalizing and solving the eigenvalue numerically, we find that the leading eigenvalue is $(k R/L_b)_1=1.003059165...$ (see also {\it S\'aez and Lecampion} [2023] and {\it Ciardo and Viesca} [2025]). Linear stability is lost when the stiffness $k=k_c$ and thus the critical radius $R_c$ is 
\begin{equation}\frac{R_c}{L_b}=\frac{1.003059165...}{1-a/b}\end{equation}
The eigenfunction corresponding to the leading eigenvalue is accurately approximated, to within an arbitrary factor, by
\begin{equation}
\mathcal{V}_1(r)=\sqrt{1-r^2}(1-r^2/2)
\end{equation}

\

For comparison, we also look to find the size of free and pinned nucleation patches that correspond to slip rate diverging in the manner of (\ref{eq:bu}). For nucleation occurring on a uniformly rate-weakening patch of the fault away from fault boundaries (Figure \ref{fig:3d}a, bottom), the nucleation is unconstrained and its shape and size must be determined. The condition $\nu=0$ and the translation and rotational symmetry of the frictional properties implies that the slip-rate distribution of (\ref{eq:bu}) will likewise be axisymmetric and translationally invariant within the rate-weakening region of the fault: $\mathcal{W}(r)$. Thus nucleation occurs over a circular region the radius of which is a free-boundary problem to be determined. The slip-rate distribution $\mathcal{W}$ and that radius $R_f$ must satisfy [{\it Viesca}, 2016a,b]
\begin{equation}
1-\frac{a}{b}+\frac{L_b}{R_f}\frac{1}{\pi}\int_{0}^1\left[\frac{E[k(r/s)]}{s-r}+\frac{K[k(r/s)]}{s+r}\right]\mathcal{W}'(s)ds=\begin{cases}1-\mathcal{W}(r) & \mathcal{W}\leq 1 \\ 0 & \mathcal{W}\geq 1\end{cases} 
\label{eq:Wr}
\end{equation}
in which distances have been normalized as $r/R_f\Rightarrow r$, $L_b=\mu D_c/\sigma b$, and the boundary condition for non-singularity of the stress rate is $\mathcal{W}'(1)=0$. This last condition determines the free-boundary problem for $R_f$. Solving this problem numerically, we find that $R_f$ may be approximated as 
\begin{equation}\frac{R_f}{L_b}\approx\begin{cases}1.726667372... & a/b<0.194448147...\\ 
\frac{\pi}{4(1-a/b)^2}\left[1+0.378(1-a/b)^2-0.642(1-a/b)^4+1.64(1-a/b)^6\right]& a/b>0.1944...\end{cases}\label{eq:Rn}\end{equation}
where the first expression in (\ref{eq:Rn}) is precise to as many significant figures shown and the second expression is accurate to within 1\% relative error up to $a/b=0.99$. The asymptotic behavior of $R_f$ as $a/b\rightarrow 1$ can be derived by the considering equivalence of the problem (\ref{eq:Rn}) to that of a slip-weakening fracture  in equilibrium with a uniform background stress [{\it Viesca}, 2016a,b]. Doing so yields that, as $a/b\rightarrow 1$,
\begin{equation}
\frac{R_f}{L_b}\sim\frac{\pi}{4(1-a/b)^2 }
\end{equation}
This nucleation length was also expressed by {\it Chen and Lapusta} [2009], citing a private communication with Rubin, presumably following the line of argument of {\it Rubin and Ampuero} [2005] based on identifying an apparent similarity between numerical solutions and a small-scale-yielding fracture. This lengthscale was re-derived by {\it Cattania and Segall} [2019] on such a basis.

\

In contrast to the free-boundary problem above, the rate-weakening patch may be circular and small enough such that nucleation, if possible, occurs within a finite, pinned radius (Figure \ref{fig:3d}a, right). We are interested in finding the smallest radius, $R_p$, that can support the spontaneous nucleation of dynamic rupture, by admitting solutions of the form (\ref{eq:bu}) satisfying (\ref{eq:Wr}). We solve (\ref{eq:Wr}) numerically by replacing $R_f$ with $R_p$ in that equation, rescaling $r$ with $R_p$ and removing the boundary condition $W’(1)=0$. In place of the latter we impose a value of $R_f/L_b$ as a problem parameter. The minimum value of that parameter that admits a numerical solution is found by bisection and is dubbed $R_p/L_b$. The dependence of $R_p$ on $a/b$ is well approximated by the expression
\begin{equation}\frac{R_p}{L_b}\,\approx\,\frac{\pi}{4(1-a/b)^2}\left[1+0.256(1-a/b)^2+0.0201(1-a/b)^4\right]\end{equation}
which is to within 0.1\% relative error for $0<a/b<0.999$.

In Figure \ref{fig:3d} we draw comparisons between the critical radius $R_c$ from linear stability and the nucleation radii $R_f$ and $R_p$. In Figure \ref{fig:3d}b we highlight that $R_p<R_f$ for $0<a/b<1$. In Figures \ref{fig:3d}c,d, we show the phase diagram of the behavior of a circular rate-weakening patch of radius $R$ driven by creep of the surrounding fault. The boundaries of the phase diagram are given by $R=R_c$ and $R=R_p$. Since $R_c<R_p$, there are three regions in the phase space, as in the equivalent problem for a single mode of slip (Figure \ref{fig:phase}). At small sizes, the circular asperity is unconditionally stable when its radius $R<R_c$, such that it can only be expected to creep steadily at the rate $V_{pl}$. At moderate sizes, $R_c<R<R_p$, the asperity is linearly unstable such that steady, uniform creep of the asperity can give way to spontaneous acceleration; however, as there is no non-linear pathway to indefinite acceleration towards dynamic rupture, such initial acceleration is ultimately limited to an aseismic transient and the fault will be expected to subsequently decelerate back towards steady state, before presumably repeating the process. For $R>R_p$, the asperity may proceed to seismic slip speeds and dynamic rupture. 

\subsection{Effect of state evolution law}
\label{sec:discse}
In section \ref{sec:ge} we presented three state evolution laws proposed by {\it Ruina} [1983]: aging, slip, and an intermediate law bridging the first two. The results from linear stability analysis apply to all three forms, as their linearizations about steady state are the same. However, as discussed in section \ref{sec:se}, the results concerning the non-linear phase of nucleation require careful consideration of the distinct non-linear forms of state evolution. The expressions for the nucleation lengths $L_f$, $L_p$, $R_f$, and $R_p$ presented in sections \ref{sec:free}, \ref{sec:pin}, and \ref{sec:3d} assumed an aging-law state evolution. Remarkably, these expressions may also be adapted to yield nucleation lengths under state-evolution laws ranging from aging to slip. As mentioned in section \ref{sec:se}, such nucleation lengths also exist and are a factor $\epsilon$ smaller than their aging-law counterparts [{\it Viesca}, 2023], where $\epsilon$ is the parameter of the intermediate state evolution law. For the aging law, $\epsilon=1$, while the slip law is retrieved in the limit $\epsilon\rightarrow 0$. In other words, for a given value of $\epsilon$ we may rewrite the free and pinned nucleation lengths as $\epsilon L_f$, $\epsilon L_p$, $\epsilon R_f$, and $\epsilon R_p$ for the single mode or mixed-mode problem configurations. 

\

\begin{figure}[t]
      \center\includegraphics[width=400pt]{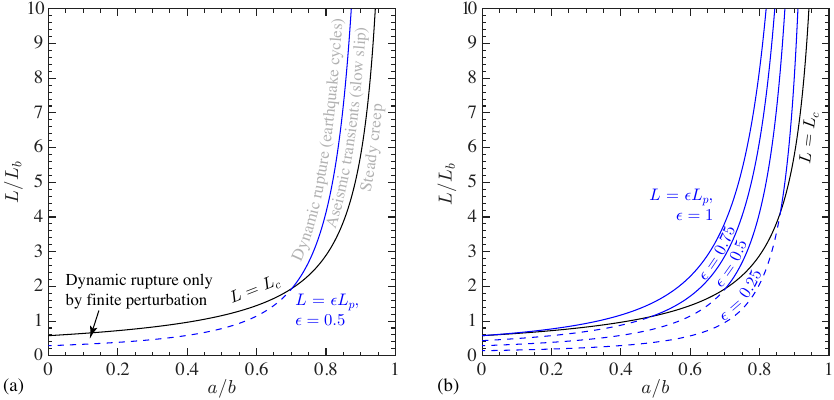}\vspace{5 mm}
   \caption{\raggedright Illustrations of the influence of the form of the state evolution law on the phase diagram delimiting regions of linear and non-linear stability. The parameter $\epsilon$ captures proximity to aging ($\epsilon=1$) and slip laws ($\epsilon\rightarrow 0$). (a) Phase diagram for expected fault behavior for a rate-weakening patch of length $2L$ within a fault that otherwise creeps at constant rate $V_{pl}$. Here the fault obeys the intermediate state evolution law with $\epsilon=0.5$. This has the effect of reducing the minimum fault length required for rupture $L_p$ by a factor $\epsilon$, such that $\epsilon L_p<L_c$ over a finite interval of $a/b$. This in turn leads to a fourth possible phase of fault behavior occurring when $\epsilon L_p<L<L_c$. In this phase the fault is linearly stable but nonlinearly unstable and finite perturbations can trigger dynamic rupture. For comparison, the equivalent aging-law phase diagram is shown in Figure \ref{fig:phase}. (b) The phase diagram shows again the boundary $L=\epsilon L_p$ drawn for several values of $\epsilon$, highlighting the increasing size of the metastable phase. }
   \label{fig:se}
\end{figure}

To highlight the consequence of this reduction in the nucleation length, we consider how the phase diagram of fault behavior changes under a state evolution other than aging ($\epsilon<1$). Specifically, we examine the case of a finite rate-weakening asperity of length $2L$ loaded by steady creep at a rate $V_{pl}$. The critical fault size $L_c$ at which linear stability is lost is given by (\ref{eq:Lc1}) and shown in Figure \ref{fig:se}a. Additionally in Figure \ref{fig:se}a, we show the minimum nucleation length $L_p$ for an aging-law fault reduced by a factor $\epsilon$ to give the correct corresponding length $\epsilon L_p$ for this intermediate evolution law. For this particular plot, we consider a state evolution law ``halfway" between aging and slip, with $\epsilon=0.5$. The consequence is the addition of a fourth possible phase of fault behavior, in addition to the previously known phases of steady creep, aseismic transients, and dynamic rupture. This fourth phase occurs when the fault length is such that $\epsilon L_p<L<L_c$. In this case, the fault is linearly stable, but non-linearly unstable. This implies that a steady uniform creep of the rate-weakening patch is stable to small perturbations, but unstable to sufficiently large perturbations: the fault will not spontaneously nucleate a dynamic rupture under the given loading conditions, but it can be driven to accelerate towards dynamic rupture in the manner of (\ref{eq:bu}) if the fault experiences a large-amplitude perturbation. In Figure \ref{fig:se}b, we highlight that  decreasing the value of $\epsilon$ towards the slip law limit ($\epsilon\rightarrow 0$) enlarges the region in phase space occupied by this fourth phase: in this limit $\epsilon L_p\rightarrow 0$, implying that any $L<L_c$ is potentially nonlinearly unstable. However, non-linear stability analyses of spring-block model with slip and intermediate-law state evolution indicate that the size of the perturbation required for instability increases with the equivalent stiffness of the system, which is expected to scale with the inverse of the fault dimension [{\it Gu et al.}, 1984; {\it Ciardo and Viesca}, 2025]: faults much smaller than $L_c$ will need a substantially larger amplitude perturbation to trigger dynamic rupture. While such spring-block models provide the best available estimate of the magnitude of the perturbation required to do so, the models have the deficiency of a single degree of freedom. The mapping of spring stiffness to fault dimension assumes the dominance of a single eigenmode, but the response of a fault in a continuum can be thought to be composed of any number of these orthogonal eigenmodes. In a linear regime these eigenmodes evolve independently, but couple under non-linearity. Thus while perturbations in spring-block models are largely characterized by an amplitude and direction in a two-dimensional state space, the additional factor of the spatial distribution of the perturbation comes into play on continuous faults. 

\

In the slip-law limit, faults are both linearly and non-linearly unstable for fault lengths $L/L_b$ greater than $L_c/L_b$. This implies that earthquakes may nucleate if this condition is met and the absence of a window in which slow slip events may spontaneously emerge, without earthquakes. This appears to be consistent with, if not completely true when comparing against numerical solutions. Specifically, numerical solutions show that the window in which slow slip events occurs is dramatically reduced, but not entirely eliminated when passing from aging to slip law. For example, {\it Rubin} [2008] considered numerical solutions with radiation damping and either aging or slip evolution laws acting on a shallowly dipping fault terminating below the surface, with a rate-weakening region up-dip and a rate-strengthening region down-dip. A slip rate was imposed at and beyond the down-dip limit of the restrengthening region and the fault was locked at the up-dip limit. He found that, for the aging law, slow-slip events  emerge above a critical rate-weakening patch size, and cease to occur when that patch size is  approximately 3.5 to 5.2 times the critical patch size, for $a/b$ values of 0.8 and 0.9, respectively. While one may repeat the linear stability analyses of section \ref{sec:linstab} for such boundary conditions and frictional property heterogeneity, for simplicity we may compare this aging-law result with that for a finite rate-weakening asperity. In this case, the comparable ratio is given by $L_p/L_c$: (\ref{eq:Lc1}) and (\ref{eq:Lp}) imply $L_p/L_c\sim [(0.578...)\pi(1-a/b)]^{-1}$ as $a/b\rightarrow 1$, yielding a ratio of 2.75 to 5.5, thus closely approximating the range of patch size in numerical solutions of {\it Rubin} [2008]. On the other hand, the {\it Rubin} [2008] numerical solutions for the slip law in the same range of $a/b$ yield slow slip events within significantly reduced range of conditions corresponding to an upper limit of $L/Lc$ between 1.5 and 1.8, but not 1 as would follow from our theoretical analysis.

\

That the window for slow slip events in these simulations diminishes, though does not vanish, may be due to the propagating nature of non-linear instabilities in slip-law simulations. Quasi-static simulations under the slip law, with $a/b>0.5$ demonstrated that late-stage acceleration occurred in the form of a propagating slip pulse [{\it Ampuero and Rubin}, 2008; {\it Rubin and Ampuero}, 2009]. Subsequently, {\it Viesca} [2023] found that solutions of the form (\ref{eq:bu}), with vanishing nucleation lengths, are unstable for $a/b>0.3781...$, meaning quasi-static acceleration remains localized when stable, but propagates when unstable. This suggests that the existence of slow-slip events for the slip law for $L>L_c$ could be the result of accelerating slip pulses being arrested at asperity boundaries, before the slip-rate can reach divergence (nucleation). In such a case, it is therefore the pulse propagation distance towards nucleation that should inform the critical nucleation size for the slip law with large enough $a/b$ [e.g., section 4.5 of {\it Rubin and Ampuero}, 2009]. This would imply that, for the slip law, nucleation is no longer a universal process, as appears to be for the aging law, though this requires further investigation.

\section{Summary and conclusion}

We examine the linear stability of faults under conditions beyond the original analysis of an unbounded fault undergoing uniform sliding.
Our analysis takes into account the finiteness of a fault and the manner in which the fault is loaded.
We find that loss of linear stability occurs as a Hopf bifurcation when the dimensionless half-length of the fault or asperity $L/L_b$ reaches a critical value, $L_c/L_b$, which is a function of the ratio of friction parameters $a/b$. This critical value diverges in the limit $a/b\rightarrow 1$ and we find its asymptotic scaling depends on the form of the steady-state initial condition. Importantly, we find that non-uniformity of the underlying steady state of sliding, due for example to slip being locally pinned at fault boundaries, results in a drastically different (larger) critical size $L_c$ compared to that for an asperity whose steady state is uniform sliding.

\

The loss of linear stability on faults or asperities with $L>L_c$ indicates the potential emergence of unstable slip transients, which may or may not reach seismic slip rates. To ascertain if the instability can eventually nucleate an earthquake or is limited to a slow-slip transient, we also consider the loss of non-linear stability in terms of non-linear solutions for diverging slip rate introduced previously in the literature [{\it Rubin and Ampuero}, 2005; {\it Viesca}, 2016a,b; {\it Viesca}, 2023].
We present an additional solution in which slip rate diverges but is spatially confined, or ``pinned", by fault boundaries.  
These pinned solutions represent the minimum nucleation length of a fault.
We provide accurate, approximate expressions for the dependence of the pinned nucleation half-length $L_p$ on the ratio $a/b$. 
We also provide such expressions for larger, so-called ``free" nucleation half-length $L_f$ as it depends on $a/b$, based on previous numerical solutions for $L_f$ over the entire range of rate-weakening $a/b$ [{\it Viesca}, 2016a,b].
$L_f$ is the expected nucleation length when the fault size is larger than $2L_f$.
Unlike the results for linear stability, $L_p$ and $L_f$ are independent of the manner of loading, with the exception of the issue of spatial confinement, and can be considered universal.

\

We show that $L_c$ and $L_p$ may be used to construct a phase diagram to determine the behavior of fault models a priori on the basis of the dimensionless parameters $L/L_c$ and $L/L_p$. If $L/L_c<1$ the system is linearly stable to perturbations. If $L/L_p<1$ the system is also nonlinearly stable, meaning that perturbations may excite a quasi-statically limited instability but no perturbations may lead to a subsequent instability that grows in a self-sustained manner without bound, other than inertial limitations. If model parameters are such that both conditions are met, the steady-state condition can be considered to be unconditionally stable. However, if $L/L_c>1$ and $L/L_p<1$ the system is linearly unstable but nonlinearly stable, meaning aseismic transients may spontaneously emerge, but cannot continue to accelerate without bound. If both $L/L_c>1$ and $L/L_p>1$, the fault is both linearly and non-linearly unstable and small perturbations can lead to a slip rate that diverges until dynamic rupture is nucleated. 

\

For conceptual simplicity, discussions in sections \ref{sec:trans} and \ref{sec:disc} were framed with fault length $L$ as a model parameter to be adjusted, with all other parameters fixed, and comparing $L$ to critical linear ($L_c$) and non-linear ($L_p$) lengths. However, one might consider the fault half-length $L$ to be fixed and we may instead consider whether an adjustment of one or more of the other problem parameters (such as $\mu$, $\nu$, $D_c$, $a$, $b$, or the effective normal stress $\sigma$) leads to a change in stability of the fault. The more general consideration to be given is that of the value of the dimensionless parameters $L/L_c$ and $L/L_p$, and specifically, whether they are greater than or less than 1. To calculate these dimensionless parameters, expressions for $L_c$ and $L_p$ in terms of $a/b$ and $L_b$ have been provided, bearing in mind that the elasto-frictional lengthscale $L_b$ is itself a function of model parameters, (\ref{eq:Lb}). 
We provide an extension of the results to 3D, identifying a critical radius $R_c$ for loss of linear stability of a circular asperity surrounded by steady creep, and the pinned/free nucleation radii $R_p$ and $R_f$. 

\

We show how the state evolution law alters the phase diagram. To do so, we use results from {\it Viesca} [2023], who considered an intermediate state evolution law of {\it Ruina} [1983], which lies on a continuum of behavior between aging to slip laws (and was also shown to be entirely equivalent to the later-proposed Nagata law [{\it Nagata et al.}, 2012; {\it Noda and Chang}, 2023; {\it Ciardo and Viesca}, 2025]).
 For the aging law, $L_c<L_p$ and we have the three possible scenarios discussed above.
 For deviations from the aging law, $L_p$ is reduced by a known factor leading to instances in the phase diagram in which $L_p<L_c$. 
 This leads to a fourth possibility that $L/L_p>1$ and $L/L_c<1$, meaning the fault is linearly stable but nonlinearly unstable. The implication of this is that the fault is susceptible to instability only under finitely large perturbations. In other words, earthquake-generating slip can only be triggered in this regime by sufficiently strong perturbations, which originate from waves, static stress changes induced by adjacent slow or seismic slip, or high-rate fluid injection.
 The window in phase space for which finite-perturbation triggering is possible grows as the state evolution law deviates from the aging law and is the largest under the slip law. 
 The linear stability results are independent of particular form of state evolution law, among those considered here. 

\appendix
\renewcommand{\theequation}{A.\arabic{equation}}
\setcounter{equation}{0}
\section{Spectral method for numerical solution of  {\it Uenishi and Rice} [2003] eigenvalue problem}
\label{app:URsoln}
Here we look to numerically solve the eigenvalue problem
\begin{equation}
\frac{1}{2\pi}\int_{-1}^1 \frac{\mathcal V'(\xi)}{x-\xi}d\xi = \eta \mathcal{V}(x)
\label{eq:URa}
\end{equation}
subject to the boundary conditions $\mathcal{V}(\pm1)=0$. Anticipating the asymptotic behavior $\mathcal{V}(x)\sim\sqrt{1\pm x}$ as $x\rightarrow\mp1$ [{\it Uenishi and Rice}, 2003], we can look for solutions for $\mathcal{V}'(x)$ in the form
\begin{equation}
\mathcal{V}'(x)=\frac{\phi(x)}{\sqrt{1-x^2}}
\label{eq:de}
\end{equation}
in which $\phi(x)$ is to be determined. Substituting (\ref{eq:de}) into the left hand side of (\ref{eq:URa}), suggests its subsequent numerical evaluation via Gaussian quadrature [e.g., {\it Boyd}, 2000], having the form
\begin{equation}
\int_{-1}^1 \frac{g(\xi,x)}{\sqrt{1-\xi^2}}d\xi\approx\sum_{j=1}^n w_j g(\xi_j,x)
\end{equation}
with weights $w_j$ and quadrature points $\xi_j$
\begin{equation}w_j=\pi/n \quad \xi_j=\cos[\pi(j-1/2)/n], \quad j=1,...,n\end{equation}
However, the presence of the singularity $(x-\xi)^{-1}$ in the integral of (\ref{eq:URa}) requires careful consideration at which points $x$ is evaluated. {\it Erdogan and Gupta} [1972] demonstrated that Gaussian quadrature continues to hold for such singular integrals provided that $x$ is evaluated at a set of points $x_i$ complementary to $\xi_j$
\begin{align}
x_i=\cos(\pi i/n) \quad \text{for} \quad i=1,...,n-1
\end{align}
such that we may approximate
\begin{equation}\frac{1}{2\pi}\int_{-1}^1 \frac{\mathcal V'(\xi)}{x_i-\xi}d\xi\approx \sum_{j=1}^nK_{ij}\phi(\xi_j) \quad i=1,...,n-1 \label{eq:Kij}\end{equation}
where
\begin{equation}K_{ij}=\frac{1}{2n}\frac{1}{x_i-\xi_j}\end{equation}

\

Now considering the right hand side of (\ref{eq:URa}), we approximate $\mathcal{V}(x_i)$ in terms of $\phi(\xi_j)$, using results from {\it Viesca and Garagash} [2018]
\begin{equation}\mathcal{V}(x_i)=\int_{-1}^{x_i} \frac{\phi(\xi)}{\sqrt{1-\xi^2}}d\xi\approx \sum_{j=1}^nS_{ij} \phi(\xi_j)\label{eq:Sij}\end{equation}
in which the matrix $S_{ij}$ can be expressed 
\begin{equation}S_{ij}=\sum_{k=0}^{n-1}A_{ik}B_{kj}\end{equation}
where
\begin{subequations}
\begin{equation} A_{ik}=\begin{cases} \pi-\arccos(x_i) & k=0 \\  -\sin[k\arccos(x_i)]/k & k\neq0 \end{cases}\end{equation}
\begin{equation} B_{kj}=\begin{cases} 1/n & k=0 \\ \displaystyle 2 \cos[k\arccos(\xi_j)]/n & k\neq0 \end{cases}\end{equation}
\end{subequations}

\

Using the approximations (\ref{eq:Kij}) and (\ref{eq:Sij}) in (\ref{eq:URa}) we arrive to the system of equations
\begin{equation}
\sum_{j=1}^nK_{ij}\phi(\xi_j)= \eta \sum_{j=1}^nS_{ij}\phi(\xi_j) \quad \quad i=1,...,n-1
\label{eq:subsys}
\end{equation}
which is nearly a discretization of the eigenvalue problem into a generalized eigenvalue problem, except that the matrices $K_{ij}$ and $S_{ij}$ are not square, but rather are $n-1$ along index $i$ by $n$ along index $j$. We add a constraint determined by the boundary conditions $\mathcal{V}(\pm1)=0$ that 
\begin{equation}\int_{-1}^1\mathcal{V}'(\xi)d\xi=0\end{equation}
the left hand side of which, via (\ref{eq:de}), may be approximated by classic Gaussian quadrature
\begin{equation}\int_{-1}^1\frac{\phi(\xi)}{\sqrt{1-\xi^2}}d\xi\approx\frac{\pi}{n}\sum_{j=1}^n\phi(\xi_j)\end{equation}
and hence the additional constraint can simply be written as
\begin{equation}\sum_{j=1}^n\phi(\xi_j)=0 \label{eq:con} \end{equation}
We may add this condition by introducing an $i=n$-th equation to the system (\ref{eq:subsys}) by updating the matrices $K_{ij}$ and $S_{ij}$ to
\begin{subequations}
\begin{equation}
\bar K_{ij} = \begin{cases} K_{ij} & i<n\\ 1 & i=n \end{cases}
\end{equation}
\begin{equation}
\bar S_{ij} = \begin{cases} S_{ij} & i<n\\ 0 & i=n \end{cases}
\label{eq:Sijbar}
\end{equation}
\label{eq:SK}
\end{subequations}
leading to the generalized eigenvalue problem
\begin{equation}\sum_{j=1}^n \bar K_{ij}\phi(\xi_j)= \eta \sum_{j=1}^n \bar S_{ij}\phi(\xi_j) \quad i=1,...,n\label{eq:geneig}\end{equation}
Solving this problem for $n$ discrete eigenvalues $\eta_q$ and $n$ discretized eigenfunctions $\phi_q(\xi_j)$ we may construct the $n$ discretized eigenfunctions $\mathcal{V}_q(x_i)$ using (\ref{eq:Sij}).

\

We may also rearrange (\ref{eq:geneig}) into a simple matrix eigenvalue problem by multiplying the left and right hand sides by $\eta^{-1}$ and the inverse of the matrix $\bar K_{ij}$, the components of which are written such that $\sum_{i=1}^n\bar K^{-1}_{ki}\bar K_{ij}=I_{jk}$ where $I_{jk}$ is the identity matrix. Consequently,
\begin{equation}\eta^{-1}\phi(\xi_k)=  \sum_{j=1}^n \bar T_{kj}\phi(\xi_j) \quad k=1,...,n\label{eq:fixeig}\end{equation}
where 
\begin{equation}\bar T_{kj}=\sum_{i=1}^n \bar K^{-1}_{ki}\bar S_{ij}\end{equation}
If we instead solve the eigenvalue problem as posed by (\ref{eq:fixeig}), we may obviously retrieve the $n$ discrete eigenvalues $\eta_q$ as the inverse of those found for $\eta^{-1}$, and we can construct the eigenfunctions using (\ref{eq:Sij}) as before. As an aside, we could not have alternatively multiplied (\ref{eq:geneig}) by the inverse of $S_{ij}$ as that inverse does not exist, which is evident considering that $S_{ij}$ is a singular matrix due to the row of zeros introduced in (\ref{eq:Sijbar}).

\section{Extension of numerical method to eigenvalue problem of linear stability with inhomogeneous base state}
\renewcommand{\theequation}{B.\arabic{equation}}
\label{app:fin}
\setcounter{equation}{0}
Here we extend the approach used to solve the {\it Uenishi and Rice} [2003] eigenvalue problem (Appendix \ref{app:URsoln}) to an eigenvalue problem that arises when considering the linear stability of a non-uniform base slip rate and state to perturbations. Specifically, we begin with the linear system of evolution equations (\ref{eq:fsys2}) repeated below
\begin{subequations}
\begin{align}
s \mathcal{V}(x)&= \frac{b}{a}\left[ V_o(x)\left(\frac{L_b}{2L}\frac{1}{\pi}\int_{-1}^1\frac{\mathcal{V}'(\xi)}{\xi-x}d\xi +\mathcal{V}(x)\right)+V^3_o(x)\Theta(x)
\right] \\[9 pt]
s\Theta(x)&=-\frac{\mathcal{V}(x)}{V_o(x)}-V_o(x)\Theta(x)
\end{align}
\label{eq:appfsys2}
\end{subequations}

We use the approximation of the integral transform, (\ref{eq:Kij}), as well as that of $\mathcal{V}(x)$, (\ref{eq:Sij}),  and discretely evaluate $\Theta(x)$ and $V(x)$ at points $x_i$ to arrive to a system of $2n-2$ equations for $2n-1$ unknowns $\phi(\xi_j)$ and $\Theta(x_i)$ 
\begin{subequations}
\begin{align}
s \sum_{j=1}^n S_{ij} \phi(\xi_j)&= \frac{b}{a}\left[ V_o(x_i)\left(-\frac{L_b}{L}\sum_{j=1}^nK_{ij}\phi(\xi_j) +\sum_{j=1}^n S_{ij}\phi(\xi_j)\right) +V^3_o(x_i)\Theta(x_i)
\right] \\[9 pt]
s\Theta(x_i)&=-\frac{1}{V_o(x_i)}\sum_{j=1}^nS_{ij}\phi(\xi_j)-V_o(x_i)\Theta(x_i)
\end{align}
\label{eq:fsysdis}
\end{subequations}
which we may more concisely express as
\begin{subequations}
\begin{align}
s \sum_{j=1}^n S_{ij} \phi(\xi_j)&= \sum_{j=1}^nM_{ij}\phi(\xi_j) +\sum_{k=1}^{n-1}N_{ik}\Theta(x_k) \quad i=1,...,n-1 \label{eq:fsysdisa} \\[9 pt]
s \sum_{k=1}^{n-1} I_{ik} \Theta(x_k)&=\sum_{j=1}^nO_{ij}\phi(\xi_j)+\sum_{k=1}^{n-1}P_{ik}\Theta(x_k)\quad i=1,...,n-1 \label{eq:fsysdisb}
\end{align}
\label{eq:fsysmat}
\end{subequations}
where $I_{ik}$ is the identity matrix and
\begin{subequations}
\begin{align}
M_{ij}&=\frac{b}{a}V_o(x_i)\left(-\frac{L_b}{L}K_{ij}+S_{ij}\right)\\[9pt]
N_{ik}&=\frac{b}{a}V_o^3(x_i)I_{ik}\\[9pt]
O_{ij}&=-\frac{1}{V_o(x_i)}S_{ij}\\[9pt]
P_{ik}&=-V_o(x_i)I_{ik}
\end{align}
\end{subequations}
As the system is underdetermined, we add the additional constraint (\ref{eq:con}),  in a similar manner as before by updating (\ref{eq:fsysdisa}) to
\begin{equation}
s \sum_{j=1}^n \bar S_{ij} \phi(\xi_j)= \sum_{j=1}^n\bar M_{ij}\phi(\xi_j) +\sum_{k=1}^{n-1}\bar N_{ik}\Theta(x_k) \quad i=1,...,n
\label{eq:fsysdisam}
\end{equation}
where $\bar S_{ij}$ is as defined in (\ref{eq:Sijbar}) and
\begin{subequations}
\begin{align}
\bar M_{ij}&=\begin{cases}\displaystyle \frac{b}{a}V_o(x_i)\left(-\frac{L_b}{L} K_{ij}+ S_{ij}\right) & i<n \\ 1 & i=n\end{cases}\\[9pt]
\bar N_{ik}&=\begin{cases}N_{ik} & i<n \\ 0 & i=n\end{cases}
\end{align}
\end{subequations}
The resulting system of $2n-1$ equations (\ref{eq:fsysdisb}) and (\ref{eq:fsysdisam}) represents a generalized eigenvalue problem for eigenvalues $s_q$ and eigenvectors $\phi_q(\xi_j)$ and $\Theta_q(x_i)$ where $q=1,...,2n-1$. 

\

As in Appendix \ref{app:URsoln}, in which we transform the generalized eigenvalue problem (\ref{eq:geneig}) to a regular one (\ref{eq:fixeig}), we may also reduce the system (\ref{eq:fsysdisb}) and (\ref{eq:fsysdisam}) to a regular eigenvalue problem. For convenience, we write the system of equations (\ref{eq:fsysdisam}) and (\ref{eq:fsysdisb}) as
\begin{equation}
s \sum_{m=1}^{2n-1} A_{im}f_m = \sum_{m=1}^{2n-1}B_{im}f_m
\label{eq:cseigp}
\end{equation}
where we have consolidated the discretized eigenfunctions into a single eigenvector
\begin{equation}
f_m=\begin{cases} \phi(\xi_m) & 1\leq m\leq n \\[9 pt] 
\Theta(x_m) & n+1\leq m \leq 2n-1\end{cases}
\end{equation}
the matrices on the left and right hand sides of (\ref{eq:fsysdisam}) and (\ref{eq:fsysdisb}) as 
\begin{equation}
A_{im}=\begin{cases} \bar S_{im} & 1\leq i\leq n, \quad 1 \leq m\leq n  \\[9 pt]
 0 & n+1\leq i \leq 2n-1,\quad 1\leq m \leq n\\[9 pt]
 0 & 1\leq i\leq n,\quad n+1\leq m\leq 2n-1 \\[9 pt]
 I_{im} & n+1\leq i \leq 2n-1, \quad n+1\leq m\leq 2n-1
 \end{cases}
\end{equation}
and
\begin{equation}
B_{im}=\begin{cases} 
\bar M_{im} & 1\leq i\leq n, \quad 1\leq m\leq n \\[9 pt] 
\bar N_{im} & 1\leq i\leq n,\quad n+1\leq m\leq 2n-1\\[9 pt] 
O_{im} & n+1\leq i \leq 2n-1,\quad 1\leq m \leq n\\[9 pt] 
P_{im} & n+1\leq i \leq 2n-1,\quad n+1 \leq m \leq 2n-1
\end{cases}
\end{equation}
We then proceed in the same manner as we did passing from (\ref{eq:geneig}) to (\ref{eq:fixeig}), multiplying both sides of (\ref{eq:cseigp}) by the inverse of the eigenvalue $s$ and the matrix $B_{im}$, leading to
\begin{equation}
\sum_{m=1}^{2n-1} C_{im}f_m = s^{-1}f_i,\quad 1\leq i\leq 2n-1
\end{equation}
with
\begin{equation} C_{im}=\sum_{k=1}^{2n-1}  B^{-1}_{ik} A_{km}\end{equation}
and $B^{-1}_{ik}$ being the components of the inverse of the matrix $B_{im}$.

\

In the eigenvalue problem of (\ref{eq:fsysdisb}) and (\ref{eq:fsysdisam}), there is a free parameter $L/L_b$. A critical value of this parameter is that for which the first unstable mode emerges. In the problems examined here, this emergence occurs as a Hopf bifurcation, when a pair of conjugate eigenvalues $s$ cross the imaginary axis leading to a positive real part, as also occurs for finite or infinite faults with a uniform sliding rate before perturbation (section \ref{sec:uni}).

\section{Eigenfunctions from linear stability analysis of problem of section \ref{sec:finuni}, when $L=L_c$}

\begin{figure}[t]
      \center\includegraphics[width=400pt]{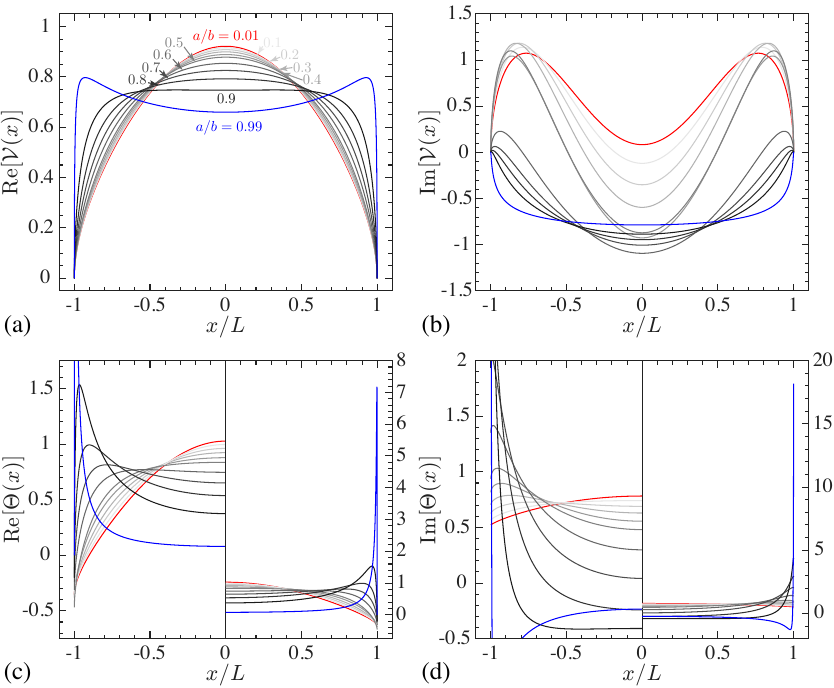}\vspace{5 mm}
   \caption{\raggedright Real and imaginary components of the eigenmodes for (a--b) velocity and (c--d) state variable perturbations, for the problem of a finite-fault loaded at constant rate of shear stress. The eigenmodes shown are those corresponding to the Hopf-bifurcating eigenmode at the critical value of $L=L_c$, at various values of $a/b$. The eigenmodes are symmetric and in panels (c) and (d), we split the display of the eigenmode $\Theta(x)$ at $x=0$ and use two different scales of the ordinate axis to highlight the divergent behavior of the eigenmode near $x=\pm 1$ at values of $a/b$ approaching 1.}
   \label{fig:fineigf}
\end{figure}

\begin{figure}[t]
      \center\includegraphics[width=400pt]{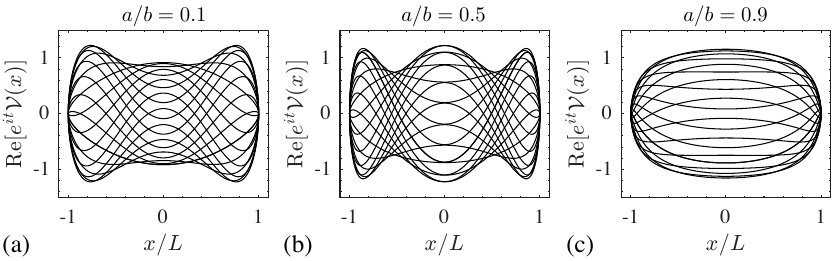}\vspace{5 mm}
   \caption{\raggedright Oscillations of the Hopf-bifurcating eigenfunction $\mathcal{V}$ of Figure \ref{fig:fineigf}, shown at fixed intervals, $\pi/10$ of the period $2\pi$ for three values of $a/b$ (a) 0.1, (b), 0.5, and (c) 0.9.}
   \label{fig:eigosc}
\end{figure}

Here we highlight a specific subset of eigenfunctions for the two-dimensional problem of a finite fault loaded at constant rate of shear stress. We consider the case for which stability is first lost: i.e., when the ratio $L/L_b$ takes on a critical value, $L_c/L_b$, the dependence of which on the ratio $a/b$ is approximated by (\ref{eq:Lc2}). At this critical value, stability is lost as a pair of complex conjugate eigenvalues crosses the imaginary axis (i.e., a Hopf bifurcation). These eigenvalues have corresponding complex eigenmodes that provide the spatial distribution of perturbations to slip rate and state: i.e., the eigenfunctions $\mathcal{V}$ and $\Theta$. In Figure \ref{fig:fineigf}, we show the real and imaginary parts of these eigenfunctions resulting from the numerical procedure highlighted in Appendix \ref{app:fin}. A given eigenfunction $g(x/L)$ is scaled such that $\int_{-1}^1 g^2(x/L)d(x/L)=1$. In darkening greyscale, we consider fixed values of $a/b$ ranging from 0.1 to 0.9 in increments of 0.1. In red and blue we highlight values of $a/b$ close to the lower and upper limits of the parameter range, respectively. We note that in the limit $a/b\rightarrow 0$, the real component of the eigenmode $\mathcal{V}$ tends towards the first eigenfunction of {\it Uenishi and Rice} [2003], approximated by (\ref{eq:eigapprox}). In Figure \ref{fig:eigosc}, we show the oscillation of the slip-rate eigenfunction $\mathcal{V}$ through a period of $2\pi$ at intervals of $\pi/10$ for three values of $a/b$: 0.1, 0.5, and 0.9.

\section{Pinned nucleation solutions}
\renewcommand{\theequation}{D.\arabic{equation}}
\label{app:pin}
\setcounter{equation}{0}
We solve for distribution $\mathcal{W}(x)$ and half-length $L_p/L_b$ that satisfy
\begin{equation}
1-\frac{a}{b}+\frac{L_b}{L_p}\frac{1}{2\pi}\int_{-1}^1 \frac{\mathcal{W}'(\xi)}{\xi-x}d\xi=\begin{cases}1-\mathcal{W}(x) & \mathcal{W}\leq 1 \\ 0 & \mathcal{W}\geq 1\end{cases} 
\label{eq:Wp}
\end{equation}
with the boundary conditions $\mathcal{W}(\pm 1)=0$. While similar to (\ref{eq:W}), here we remove the condition precluding a singularity that determined the so-called ``free" nucleation length $L_f/L_b$ in (\ref{eq:W}). To look for so-called ``pinned" nucleation solutions, we discretize (\ref{eq:Wp}) according to a Gauss-Chebyshev quadrature admitting endpoint singularity [{\it Viesca and Garagash}, 2018] and numerically solve the resulting system of equations with Newton-Raphson iteration for the discretized $\mathcal{W}$ for a given value of $L_p/L_b$. The procedure for comparable problems is discussed in detail by {\it Viesca and Garagash} [2018].

\begin{figure}[t]
      \center\includegraphics[width=400pt]{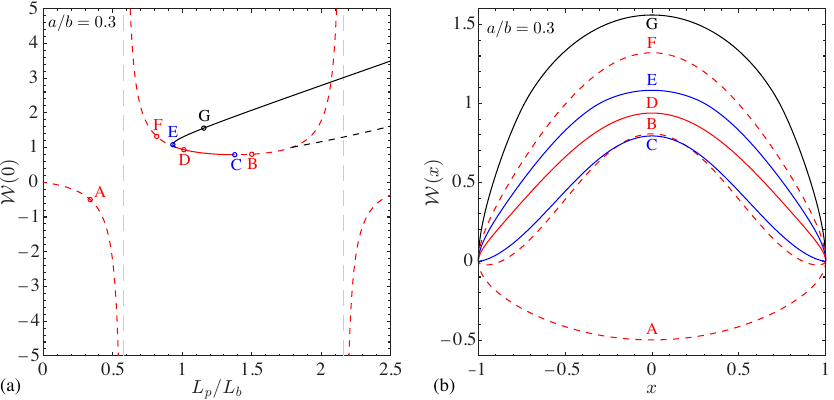}\vspace{5 mm}
   \caption{\raggedright
Examination of families of non-linear instability solutions of the form (\ref{eq:bu}), illustrated using the relation between the central value of the normalized slip rate, $\mathcal{W}(0)$, and the size of their domain of support $|x|<L_p$, for the particular case of $a/b=0.3$. Solutions in red are solutions to (\ref{eq:W}) with the righthand side replaced with $1-\mathcal{W}$. Red-dashed solutions are inadmissible either because $\mathcal{W}$ takes on negative values (unphysical) or exceeds 1 (and hence not a solution to (\ref{eq:Wp})). Solid red are admissible solutions for which $0<\mathcal{W}\leq 1$. Solutions in solid black are admissible solutions to (\ref{eq:Wp}) for which $\mathcal{W}(0)>1$ and in dashed-black are inadmissible solutions for which $\mathcal{W}$ takes on negative values. (a) Plot of solutions branches of $\mathcal{W}(0)$ vs. $L_p/L_b$. Grey-dashed lines highlight solution asymptotes at $L/L_b=0.578...$ and $2.15...$ . (b) Plot of particular solutions of $\mathcal{W}(x)$ corresponding to points A--G in (a). The blue curves of C and E correspond, respectively, to the free nucleation solution with $L_f/L_b$ and the smallest pinned nucleation solution with $L_p^{min}/L_b$ (simply referred to as $L_p/L_b$ in the main text).}
   \label{fig:pinapp}
\end{figure}

\

The solution of problem (\ref{eq:Wp}), illustrated in Figure \ref{fig:pinapp}, exists and is physical ($\mathcal{W}$ non-negative) only for $L_p>L_{p}^{min}>0.5788...$. At a given value of $a/b$, the physical, non-negative solution of (\ref{eq:Wp}) is non-unique and is shown by the solid red ($\mathcal{W}(0)\leq 1$) and solid black ($\mathcal{W}(0)>1$) branches. The fault size $L_p$ along the lower solid solution branch is upper-bounded by $L_p=L_p^{(C)}=L_f$ (point $C$, Figure \ref{fig:pinapp}a), corresponding to the ``free" solution (with vanishing tip singularity, $\mathcal{W}'(\pm 1)=0$, Figure \ref{fig:pinapp}b) and lower-bounded by $L_p=L_p^{(E)}=L_p^{min}$ (point E, Figure \ref{fig:pinapp}a). The lower solution branch therefore corresponds to fault size window $L_p^{min}\leq L_p\leq L_f$. A solution does not exist for $L_p<L_p^{min}$ and is nonphysical ($\mathcal{W}<0$ near the pinned boundaries) for $L_p>L_f$. The upper solution branch in black is continuous with the lower red branch and the two join near $L_p=L_p^{min}$. The upper black branch corresponds to increasing $L_p$ and $\mathcal{W}(0)$, Figure \ref{fig:pinapp}a. This therefore defines two nucleation-relevant sizes, $L_p^{min}$ (singular or ``pinned" nucleation) and $L_f$ (non-singular, or ``free" nucleation).

\

To understand the likely origin of the solution non-uniqueness and non-existence below $L_p^{min}$, we can consider a complementary problem given by (\ref{eq:Wp}) with the right hand side replaced by $1-\mathcal{W}$, now unconditionally of the value of $\mathcal{W}$. Such a problem is linear and the solution is unique and shown by the red-dashed line in Figure \ref{fig:pinapp}a. When $L_p$ is varied, the solution for $\mathcal{W}(0)$ alternates between positive and negative values crossing the even-mode eigenvalues of the {\it Uenishi and Rice} [2003] eigenvalue problem (0.57888694..., 2.1584005..., etc.), highlighted by the grey-dashed asymptotes, at which point the solution diverges. Within these intervals, local minima and maxima occur at odd-mode eigenvalues of {\it Uenishi and Rice} [2003] (1.3773773..., 2.9460737...). In Figure \ref{fig:pinapp}, a point along the solid branch (in the vicinity of point E) marks the transition from solid red to solid black: at this point $\mathcal{W}(0)=1$ and thus the departure of the non-linear problem solution (solid line) from the complementary linear one (red-dashed) as $L_p$ is decreased below this point. The segment between this point and point $C$ is the solid red branch where $0<\mathcal{W}(x)\leq1$ for all $x$ and the non-linear problem (\ref{eq:Wp}) coincides with the linear one. Beyond point $C$, with $L>L_f$, solutions of both linear (red-dashed) and non-linear (black dashed) problems continue, but lose physical meaning as $\mathcal{W}<0$ near the pinned boundaries. 

\

In Figure \ref{fig:pinapp}b we plot select solutions for $\mathcal{W}(x)$ from among the two solution families of Figure \ref{fig:pinapp}a. We plot the specific cases of Figure \ref{fig:pinapp}a indicated by circles, the colors of which correspond to line colors in Figure \ref{fig:pinapp}b. As in Figure \ref{fig:pinapp}, solid lines correspond to physically admissible solutions to (\ref{eq:Wp}) (cases C, D, E, and G) while dashed lines are inadmissible solutions either because they are unphysical solutions to (\ref{eq:Wp}) owing to $\mathcal{W}$ having negative values (cases A and B) or because they are solutions to a different but similar problem of (\ref{eq:Wp}) for which the right hand side is strictly $1-\mathcal{W}$ and for which $\mathcal{W}$ exceeds 1 (case F), and hence does not respect the original problem (\ref{eq:Wp}). The two blue curves C and E do not respect the black/red color convention but are colored as such to highlight their singular nature of being the ``free/unpinned" nucleation solution and the smallest pinned nucleation solution, respectively. $L_p^{min}$ (referred to simply as $L_p$ in the main text) is thus obtained at discrete values of $a/b$ in the range $0<a/b<1$ to arrive to the solutions plotted in Figure \ref{fig:nonlin} and in subsequent figures. A similar procedure is used to arrive to the minimum pinned nucleation radius $R_p$ for a confined circular rate-weakening patch, shown in Figure \ref{fig:3d}.

There are two peculiarities worth further discussion and analysis: pinned solutions are non-unique for $L_p^{min}<L_p\leq L_f$ and an admissible branch of pinned solutions continues to exist for $L_p>L_f$. When solutions are non-unique, we cannot yet say which of the two would provide the likely path for slip rate divergence in the non-linear regime. A complete analysis would examine the asymptotic stability of each solution [{\it Viesca}, 2016a,b]. Each solution may be asymptotically stable (all eigenvalues have negative real part) or attractive (all eigenvalues have negative real part, save a few unstable modes) or unattractive (all eigenvalues have positive real part). Performing the analysis and presuming the solutions are at least attractive, we may find that one of the two solutions is more attractive and, hence, likely dictates the rate and spatial distribution with which slip rate diverges. However, in this case, the existence of attractive solutions is paramount and non-uniqueness is a secondary consideration for the practical purpose of determining fault or asperity sizes capable of nucleating dynamic rupture: we identify a continuous existence of pinned solutions $L_p$ below $L_f$ down to a minimum value $L^{min}_p$. 

\

Turning to the second peculiarity, solutions exist with $L_p>L_f$ (i.e., beginning on the solid black curve above point C and continuing to the right). For a fault large enough to accommodate such a pinned nucleation with half-length $L_p$ there remains an alternative pathway for instability: unpinned or ``free" nucleation over a length $L_f<L_p$ within the fault or asperity. While we do not perform a comparative analysis of the two solutions' asymptotic stability and relative attractiveness, it seems reasonable to presume that nucleation would occur over the smaller of the two nucleation half-lengths ($L_f$). 

\

As an aside, we can provide an asymptotic solution for $\mathcal{W}(x)$ along this branch in the limit of large $L_p/L_b$. In this limit, $\mathcal{W}<1$ only within a diminishing region near $x=\pm 1$, and $\mathcal{W}>1$ over nearly the entire domain $|x|<1$ such that $\mathcal{W}$ satisfies a simpler version of (\ref{eq:Wp})
\begin{equation}
1-\frac{a}{b}+\frac{L_b}{L_p}\frac{1}{2\pi}\int_{-1}^1 \frac{\mathcal{W}'(\xi)}{\xi-x}d\xi=0 
\label{eq:Wp1}
\end{equation}
The solution to which is $\mathcal{W}(x)=2(1-a/b)(L_p/L_b)\sqrt{1-x^2}$, yielding the asymptotic scaling of the solid black branch of Figure \ref{fig:pinapp}a at large $L_p/L_b$: $\mathcal{W}(0)=2(1-a/b)L_p/L_b$ (not shown in figure). This resembles a small-scale-yielding solution, except here a singularity is apparent at all scales, both on distances comparable to $x$ and those within the small boundary-layer for which $\mathcal{W}<1$. For comparison, we note that the problem for $L_f$ in the limit $a/b\rightarrow 1$ reduces to a small-scale yielding problem for which a singularity appears to be present on distances comparable to $x$, but is not present when examining distances comparable to cohesive-zone boundary layers near the tips. For this unpinned/``free" nucleation solution, $\mathcal{W}(x)\sim 2(1-a/b)(L_f/L_b)\sqrt{1-x^2}$, where $L_f/L_b\sim (1-a/b)^{-2}/\pi$. The asymptotic matching of the pinned and unpinned solutions in the limit $a/b\rightarrow 1$ accounts for the asymptotic approach of $L_p$ to $L_f$ in that limit.

\

\begin{figure}
      \center\includegraphics[width=400pt]{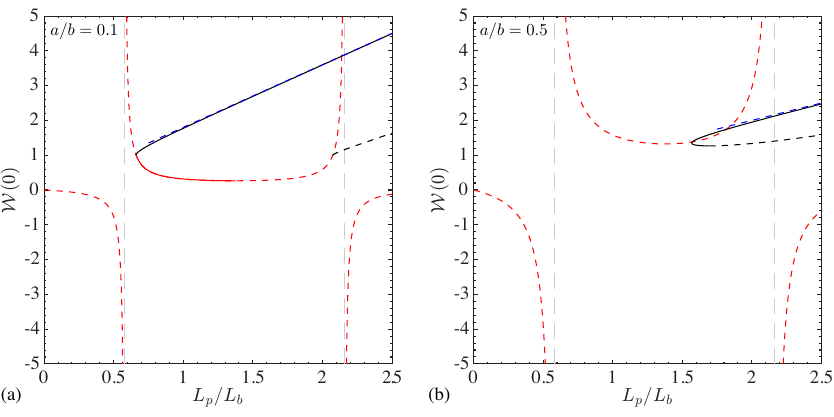}\vspace{5 mm}
   \caption{\raggedright
Same as Figure \ref{fig:pinapp}a except for (a) $a/b=0.1$ and (b) $a/b=0.5$. Blue-dashed lines are the asymptotic behavior for large $L_p/L_b$ provided by the solution to (\ref{eq:Wp1}).}
   \label{fig:pinapp2}
\end{figure}

In Figure \ref{fig:pinapp2} we present a plot in the manner of Figure \ref{fig:pinapp}a, except with $a/b$ changed to $a/b=0.1$ (left) and $a/b=0.5$ (right). In the former case we see the approach of $L^{min}_p/L_b$ towards the vertical asymptote at $L_p/L_b=0.578...$, its asymptotic limit as $a/b\rightarrow 0$. In the latter case all of the admissible solutions to (\ref{eq:Wp}) contain $\mathcal{W}>1$, such that none of the solutions to the reduced problem represented by the red-dashed-line is a solution to the full problem (\ref{eq:Wp}). 

\

\noindent{\bf References}

\medskip

\noindent Aldam, M., Y. Bar-Sinai, I. Svetlizky, E. A. Brener, J. Fineberg, and E. Bouchbinder (2016) Frictional sliding without geometric reflection symmetry, {\it Phys. Rev. X}, 6, 041023. doi:PhysRevX.6.041023

\medskip

\noindent Ampuero, J.-P., and A. M. Rubin (2008) Earthquake nucleation on rate and state faults — aging and slip laws, {\it J. Geophys. Res.}, 113, B01302, doi:10.1029/2007JB005082

\medskip

\noindent Barbot, S. (2019) Slow-slip, slow earthquakes, period-two cycles, full and partial ruptures, and deterministic chaos in a single asperity fault, {\it Tectonophys.}, 768, 228171.\\ \noindent doi:10.1016/j.tecto.2019.228171

\medskip

\noindent Ben-Zion, Y., and J. R. Rice (1995) Slip patterns and earthquake populations along different classes of faults in elastic solids, {\it J. Geophys. Res.}, 100(B7), 12,959--12983.\\ \noindent doi:10.1029/94JB03037

\medskip

\noindent Bhattacharya, P. and R. C. Viesca (2019) Fluid-induced aseismic fault slip outpaces pore-fluid migration, {\it Science}, 364 (6439), 464--468. doi:10.1126/science.aaw7354

\medskip

\noindent Blanpied, M. L., D. A. Lockner, and J. D. Byerlee (1991) Fault stability inferred from granite sliding experiments at hydrothermal conditions,
{\it Geophys. Res. Lett.}, 18(4), 609--612. doi:10.1029/91GL00469

\medskip

\noindent Blanpied, M. L., D. A. Lockner, and J. D. Byerlee (1995) Frictional slip of granite at hydrothermal conditions, {\it J. Geophys. Res.},
100(B7), 13,045--13,064. doi:10.1029/95JB00862

\medskip

\noindent Brener, E. A., M. Weikamp, R. Spatschek, Y. Bar-Sinai, and E. Bouchbinder (2016) Dynamic instabilities of frictional sliding at a bimaterial interface, {\it J. Mech. Phys. Solids}, 89, 149--173. doi:10.1016/j.jmps.2016.01.009
\medskip

\noindent Boyd, J. P. (2000) Chebyshev and Fourier Spectral Methods, {\it Dover}, Mineola, NY.

\medskip

\noindent Cao, T., and K. Aki (1986) Seismicity simulation with a rate- and state-dependent friction law, {\it Pure Appl. Geophys.} 124(3), 488--513. doi:10.1007/BF00877213

\medskip

\noindent Cattania, C., and P. Segall (2019) Crack models of repeating earthquakes predict observed moment-recurrence scaling, {\it J. Geophys. Res.}, 124, 476--504. doi:10.1029/2018JB016056

\medskip

\noindent Chen, T., and N. Lapusta (2009) Scaling of small repeating earthquakes explained by interaction of seismic and aseismic slip in a rate and state fault model, {\it J. Geophys. Res.}, 114, B01311, doi:10.1029/2008JB005749

\medskip

\noindent Chen, T., and N. Lapusta (2019) On behaviour and scaling of small repeating earthquakes in rate and state fault models, {\it Geophys. J. Int.}, 218(3), 2001--2018, doi:10.1093/gji/ggz270

\medskip

\noindent Ciardo, F., and R. C. Viesca (2025) Nonlinear stability analysis of slip in a single degree of freedom elastic system with frictional evolution laws spanning ageing to slip, {\it Proc. Roy. Soc. A}, 481, 20240917. doi:10.1098/rspa.2024.0917.

\medskip

\noindent Dieterich, J. H. (1978) Time-dependent friction and the mechanics of stick-slip, {\it Pure Appl. Geophys.}, 116(4), 790--806. doi:10.1007/BF00876539

\medskip

\noindent Dieterich, J. H. (1979) Modeling of Rock Friction 1. Experimental Results and Constitutive Equations, {\it J. Geophys. Res.}, 84(B5), 2161--2168. doi:10.1029/JB084iB05p02161

\medskip

\noindent Dieterich, J. H. (1986) A model for the nucleation of earthquake slip, in {\it Earthquake Source Mechanics}, Geophysical Monograph Series, vol. 37, eds. S. Das, J. Boatwright, and C. H. Scholz, pp. 37--47., American Geophysical Union. doi:10.1029/GM037p0037

\medskip

\noindent Dragert, H., K. Wang, T. S. James (2001) A silent slip event on the deeper Cascadia subduction interface, {\it Science}, 292(5521), 1525--1528. doi:10.1126/science.1060152

\medskip

\noindent Erdogan, F., and G. D. Gupta (1972), On the numerical solution of singular integral equations, {\it Q. Appl. Math.}, 29, 525--534. doi:10.1090/qam/408277

\medskip

\noindent Gao, H. (1988) Nearly circular shear mode cracks, {\it Int. J. Solids Struct.}, 24(2), 177--193. doi:10.1016/0020-7683(88)90028-5

\medskip

\noindent Gu, J. C., J. R. Rice, A. L. Ruina, and S. T. Tse (1984) Slip motion and stability of a single degree of freedom elastic system with rate and state dependent friction, {\it J.Mech. Phys. Solids}, 32, 167--196. doi:10.1016/0022-5096(84)90007-3

\medskip

\noindent Heimisson, E. R., E. M. Dunham, and M. Almquist (2019) Poroelastic effects destabilize mildly rate-strengthening friction to generate stable slow slip pulses, {\it J. Mech. Phys. Solids}, 130, 262--279. doi:10.1016/j.jmps.2019.06.007

\medskip

\noindent Heimisson, E. R., J. Rudnicki, and N. Lapusta (2021) Dilatancy and compaction of a rate-and-state fault in a poroelastic medium: Linearized stability analysis, {\it J. Geophys. Res.}, 126, e2021JB022071. doi:10.1029/2021JB022071

\medskip

\noindent Hirose, H., K. Hirahara, F. Kimata, N. Fujii, and S. Miyazaki (1999) A slow thrust slip event following the two 1996 Hyuganada Earthquakes beneath the Bungo Channel, southwest Japan, {\it Geophys. Res. Lett.}, 26(21), 3237--3240. doi:10.1029/1999GL010999

\medskip

\noindent Kato, N., and T. Hirasawa (1997) A numerical study on seismic coupling along subduction zones using a laboratory-derived friction law, {\it  Phys. Earth Planet. In.}, 102, 51--68. doi:10.1016/S0031-9201(96)03264-5

\medskip

\noindent Kato, N. (2003) Repeating slip events at a circular asperity: numerical simulation with a rate- and state-dependent friction law, {\it Bull. Earthq. Res. Inst. Univ. Tokyo}, 78(3), 151--166. 

\medskip

\noindent King, F. W. (2009) Hilbert Transforms, vol. 1, Cambridge University Press, Cambridge, UK. doi:10.1017/CBO9780511721458

\medskip

\noindent Lambert, V., and N. Lapusta (2021) Resolving simulated sequences of earthquakes and fault interactions: implications for physics-based seismic hazard assessment, {\it J. Geophys. Res.}, 126, e2021JB022193, doi:10.1029/2021JB022193.

\medskip

\noindent Lapusta, N., J. R. Rice, Y. Ben-Zion, G. Zheng (2000) Elastodynamic analysis for slow tectonic loading with spontaneous rupture episodes on faults with rate- and state-dependent friction, {\it J. Geophys. Res.}, 105(B10), 23,765--23,789. doi:10.1029/2000JB900250

\medskip

\noindent Lapusta, N., and Y. Liu (2009) Three-dimensional boundary integral modeling of spontaneous earthquake sequences and aseismic slip, {\it J. Geophys. Res.}, 114, B09303.\\ \noindent doi:10.1029/2008JB005934

\medskip

\noindent Lebihain, M., T. Roch, M. Violay, and J.-F. Molinari (2021) Earthquake nucleation on faults with heterogeneous weakening rate, {\it Geophys. Res. Lett.}, 48, e2021GL094901.\\ \noindent doi:10.1029/2021GL094901

\medskip

\noindent Liu, Y., and J. R. Rice (2005) Aseismic slip transients emerge spontaneously in three-dimensional rate and state modeling of subduction earthquake sequences, {\it J. Geophys. Res.}, 110, B08307. doi:10.1029/2004JB003424

\medskip

\noindent Liu, Y., and J. R. Rice (2007) Spontaneous and triggered aseismic deformation transients in a subduction fault model, {\it J. Geophys. Res.}, 112, B09404. doi:10.1029/2007JB004930

\medskip

\noindent Luo, Y., and J.-P. Ampuero (2018) Stability of faults with heterogeneous friction properties and effective normal stress, {\it Tectonophys.}, 733, 257--272. doi:10.1016/j.tecto.2017.11.006

\medskip

\noindent Mei, C. (2025) Slip instability of dilatant and fluid-saturated faults, {\it Tectonophys.}, 895, 230598. doi:10.1016/j.tecto.2024.230598

\medskip

\noindent Nagata, K., M. Nakatani, and S. Yoshida (2012) A revised rate- and state-dependent friction law obtained by constraining constitutive and evolution laws separately with laboratory data, {\it J. Geophys. Res.}, 117, B02314. doi:10.1029/2011JB008818

\medskip

\noindent Nie, S., and S. Barbot (2021) Seismogenic and tremorgenic slow slip near the stability transition of frictional sliding, {\it Earth Planet. Sci. Lett.}, 569, 117037. doi:10.1016/j.epsl.2021.117037

\medskip

\noindent Noda, H., and C. Chang, (2023) Tertiary creep behavior for various rate- and state-dependent friction laws, {\it Earth Planet. Sci. Lett.}, 619, 118314. doi:10.1016/j.epsl.2023.118314

\medskip

\noindent Noda, H., and M. Yamamoto, (2024) Homoclinic bifurcation of a rate-weakening patch in a viscoelastic medium and effect of strength contrast, {\it Earth Planets Space}, 76, 155. doi:10.1186/s40623-024-02100-w

\medskip

\noindent Ozawa, S., Y. Yang, and E. M. Dunham (2024) Fault-valve instability: A mechanism for slow slip events, {\it J. Geophys. Res.}, 129, e2024JB029165. doi:10.1029/2024JB029165
\medskip

\medskip

\noindent Ray, S., and R. C. Viesca (2017) Earthquake nucleation on faults with heterogeneous frictional properties, normal stress, {\it J. Geophys. Res.}, 122, 8214–8240. doi:10.1002/2017JB014521

\medskip

\noindent Ray, S., and R. C. Viesca (2019) Homogenization of fault frictional properties, {\it Geophys. J. Int.}, 219, 1203--1211. doi:10.1093/gji/ggz327

\medskip 

\noindent Rice, J. R. (1968) Mathematical analysis in the mechanics of fracture, in {\it Fracture, an Advanced Treatise}, vol. II, edited by H. Liebowitz, chap. 3, pp. 191--311, Academic, New York.

\medskip 

\noindent Rice, J. R. (1985) First-order variation in elastic fields due to variation in location of a planar crack front, {\it J. Appl. Mech.}, 52(3), 571--579. doi:10.1115/1.3169103

\medskip

\noindent Rice, J. R., and S. T. Tse (1986) Dynamic motion of a single degree of freedom system following a rate and state dependent friction law, {\it J. Geophys. Res.}, 91(B1), 521--530.\\ \noindent doi:10.1029/JB091iB01p00521

\medskip

\noindent Rice, J. R. (1993) Spatio-temporal complexity of slip on a fault, {\it J. Geophys. Res.}, 98(B6), 9885--9907. doi:10.1029/93JB00191

\medskip

\noindent Rice, J. R., N. Lapusta, and K. Ranjith (2001) Rate and state dependent friction and the stability of sliding between elastically deformable solids, {\it J. Mech. Phys. Solids}, 49, 1865--1898. doi:10.1016/S0022-5096(01)00042-4

\medskip

\noindent Rubin, A. M., and J.-P. Ampuero (2005) Earthquake nucleation on (aging) rate and state faults, {\it J. Geophys. Res.}, 110, B11312. doi:10.1029/2005JB003686

\medskip

\noindent Rubin, A. M. (2008) Episodic slow slip events and rate-and-state friction, {\it J. Geophys. Res.}, 113, B11414. doi:10.1029/2008JB005642

\medskip 

\noindent Rubin, A. M., and J.-P. Ampuero (2009) Self-similar slip pulses during rate-and-state earthquake nucleation, {\it J. Geophys. Res.}, 114, B11304. doi:10.1029/2009JB006529.

\medskip

\noindent Rice, J. R., and A. L. Ruina (1983) Stability of steady frictional slipping, {\it J. Appl. Mech.}, 105, 343--349. doi:10.1115/1.3167042

\medskip 
\noindent Rice, J. R., and A. L. Ruina (1983) Stability of steady frictional slipping, {\it J. Appl. Mech.}, 105, 343--349. doi:10.1115/1.3167042

\medskip 
\noindent S\'aez, A., B. Lecampion, P. Bhattacharya, and R. C. Viesca (2022) Three-dimensional fluid-driven stable frictional ruptures, {\it J. Mech. Phys. Solids}, 160, 104754.\\ \noindent doi:10.1016/j.jmps.2021.104754

\medskip

\noindent S\'aez, A., and B. Lecampion (2023) Fluid-driven slow slip and earthquake nucleation on a slip-weakening circular fault, {\it J. Mech. Phys. Solids}, 183, 105506. \\ \noindent doi:10.1016/j.jmps.2023.105506

\medskip

\noindent S\'aez, A., F. Passelègue, and B. Lecampion (2025) Maximum size and magnitude of injection-­induced slow slip events, {\it Sci. Adv.}, 11, eadq0662. \noindent doi:10.1126/sciadv.adq0662

\medskip

\noindent Salamon, N., and J. Dundurs (1971) Elastic fields of a dislocation loop in a two-phase material, {\it J. Elasticity}, 1 (2), 153--164. doi:10.1007/BF00046466

\medskip

\noindent Salamon, N., and J. Dundurs (1977) A circular glide dislocation loop in a two-phase material, {\it J. Phys. C: Solid State Phys.}, 10 (4), 497--507. doi:10.1088/0022-3719/10/4/007

\medskip

\noindent Segall, P., and J. R. Rice (1995) Dilatancy, compaction, and slip-instability of a fluid-infiltrated fault, {\it J. Geophys. Res.}, 100(B11), 22,155--22,171. doi:10.1029/95JB02403
\medskip

\noindent Segall, P. (2010) Earthquake and volcano deformation, Princeton University Press, Princeton, NJ. doi: 10.1515/9781400833856

\medskip

\noindent Shibazaki, B., and T. Shimamoto (2007) Modelling of short-interval silent slip events in deeper subduction interfaces considering the frictional properties at the unstable-stable transition regime, {\it Geophys. J. Int.}, 171(1), 191--205. doi:10.1111/j.1365-246X.2007.03434.x

\medskip

\noindent Stesky, R. M. (1975) The mechanical behavior of faulted rock at high temperature and pressure, Ph.D. thesis, M. I. T., Cambridge. 

\medskip

\noindent Tse, S. T., and J. R. Rice (1986) Crustal earthquake instability in relation to the depth variation of frictional slip properties, {\it J. Geophys. Res.}, 91(B9), 9452--9472.\\ \noindent doi:10.1029/JB091iB09p09452

\medskip

\noindent Uenishi, K., and J. R. Rice (2003), Universal nucleation length for slip-weakening rupture instability under nonuniform fault loading, {\it J. Geophys. Res.}, 108(B1), 2043.\\ \noindent doi:10.1029/2001JB001681

\medskip

\noindent Veedu, D. M., C. Giorgetti, M. Scuderi, S. Barbot, C. Marone, and C. Collettini (2020) Bifurcations at the stability transition of earthquake faulting, {\it Geophys. Res. Lett.}, 47, e2020GL087985. doi:10.1029/2020GL087985

\medskip

\noindent Viesca, R. C. (2016a) Stable and unstable development of an interfacial sliding instability, {\it Phys. Rev. E}, 93, 060202(R). doi:10.1103/PhysRevE.93.060202

\medskip

\noindent Viesca, R. C. (2016b) Self-similar slip instability on interfaces with rate- and state-dependent friction, {\it Proc. Roy. Soc. A}, 472, 20160254. doi:10.1098/rspa.2016.0254

\medskip

\noindent Viesca, R. C., and D. I. Garagash (2018) Numerical methods for coupled fracture problems, {\it J. Mech. Phys. Solids}, 113, 13--34. doi:10.1016/j.jmps.2018.01.008

\medskip

\noindent Viesca, R. C., and P. Dublanchet (2019) The slow slip of viscous faults, {\it J. Geophys. Res.}, 124, 4959--4983. doi:10.1029/2018JB016294

\medskip

\noindent Viesca, R. C. (2023) Frictional state evolution laws and the non-linear nucleation of dynamic shear rupture, {\it J. Mech. Phys. Solids}, 173, 105221. doi:10.1016/j.jmps.2023.105221

\medskip

\noindent Viesca, R. C. (2025) Asymptotic solutions for self-similarly expanding fault slip induced by fluid injection at constant rate, {\it J. Fluid Mech.}, 1019, A27, doi:10.1017/jfm.2025.10593

\medskip

\noindent Yabe, S., and S. Ide (2017) Slip-behavior transitions of a heterogeneous linear fault, {\it J. Geophys. Res.}, 122, 387--410, doi:10.1002/2016JB013132

\medskip

\noindent Zhai, P., and Y. Huang (2024) The effects of characteristic weakening distance on earthquake nucleation styles in fully dynamic seismic cycle simulations., {\it J. Geophys. Res.}, 129, e2024JB029719, doi:10.1029/2024JB029719

\end{document}